\documentclass[fleqn,usenatbib]{mnras}

\usepackage[T1]{fontenc}

\DeclareRobustCommand{\VAN}[3]{#2}
\let\VANthebibliography\thebibliography
\def\thebibliography{\DeclareRobustCommand{\VAN}[3]{##3}\VANthebibliography}

\usepackage{graphicx}	
\usepackage{amsmath}	
\usepackage{amssymb}	

\usepackage{orcidlink}
\usepackage{newtxtext,newtxmath}

\title[DM Annihilation Signals From the Sagittarius Dwarf]{Prospective Dark Matter Annihilation Signals From the Sagittarius Dwarf Spheroidal}
\author[T. A. A. Venville et. al.]{Thomas A. A. Venville,$^{\orcidlink{0000-0003-0278-9933}}$$^{1,2}$\thanks{101615311@student.swin.edu.au}
Alan R. Duffy,$^{1,2}$
Roland M. Crocker$^{\orcidlink{0000-0002-2036-2426}}$$^{3}$,
Oscar Macias$^{\orcidlink{0000-0001-8867-2693}}$$^{4,5}$
\newauthor and Thor Tepper-García$^{\orcidlink{0000-0002-1081-883X}}$$^{6,7}$
\\
$^{1}$Centre for Astrophysics and Supercomputing, Swinburne University of Technology, PO Box 218, Hawthorn, Victoria, 3122, Australia \\
$^{2}$ARC Centre of Excellence for Dark Matter Particle Physics, Australia \\
$^{3}$Research School of Astronomy and Astrophysics, Australian National University, Canberra 2611, A.C.T., Australia \\
$^{4}$Department of Physics and Astronomy, San Francisco State University, San Francisco, CA 94132, USA \\
$^{5}$GRAPPA $-$ Gravitational and Astroparticle Physics Amsterdam, University of Amsterdam, Science Park 904, 1098 XH Amsterdam, The Netherlands\\ 
$^{6}$Sydney Institute for Astronomy, School of Physics, The University of Sydney, NSW 2006, Australia\\
$^{7}$Centre of Excellence for All Sky Astrophysics in Three Dimensions (ASTRO-3D), Australia
}
\date{Accepted XXX. Received YYY; in original form ZZZ}

\pubyear{2023}

\begin{document}
\label{firstpage}
\pagerange{\pageref{firstpage}--\pageref{lastpage}}
\maketitle

\begin{abstract}
The Sagittarius Dwarf Spheroidal galaxy (Sgr) is investigated as a target for DM annihilation searches utilising J-factor distributions calculated directly from a high-resolution hydrodynamic simulation of the infall and tidal disruption of Sgr around the Milky Way. In contrast to past studies, the simulation incorporates DM, stellar and gaseous components for both the Milky Way and the Sgr progenitor galaxy. The simulated distributions account for significant tidal disruption affecting the DM density profile. Our estimate of the J-factor value for Sgr, $J_{\text{Sgr}}=1.48\times 10^{10}$ M$_\odot^2$ kpc$^{-5}$ ($6.46\times10^{16}\ \text{GeV}\ \text{cm}^{-5}$), is significantly lower than found in prior studies. This value, while formally a lower limit, is likely close to the true J-factor value for Sgr. It implies a DM cross-section incompatibly large in comparison with existing constraints would be required to attribute recently observed $\gamma$-ray emission from Sgr to DM annihilation. We also calculate a J-factor value using a NFW profile fitted to the simulated DM density distribution to facilitate comparison with past studies. This NFW J-factor value supports the conclusion that most past studies have overestimated the dark matter density of Sgr on small scales. This, together with the fact that the Sgr has recently been shown to emit $\gamma$-rays of astrophysical origin, complicate the use of Sgr in indirect DM detection searches.
\end{abstract}


\section{Introduction: dark matter annihilation signals from dwarf spheroidal Galaxies}
\label{sec:Introduction}
In the $\Lambda \text{CDM}$ cosmological model of the Universe, approximately $83\%$ of the total mass density of the universe consists of dark matter (DM), 
a massive particle species that primarily interacts with baryonic matter through gravitational interactions \citep{Garrett_2011}. The hierarchical gravitational formation of structure in this cosmological model results in galaxies contained in more massive DM haloes. These DM haloes are thus often traced by stellar populations. Diverse experiments have attempted to detect particle DM candidates, targeting a wide range of DM masses and velocity averaged annihilation cross sections \citep[e.g.][]{Bertone_2005}. These experimental searches include monitoring for direct detection of DM interaction with target materials and `indirect' searches for Standard Model products of DM self-annihilation and decay, for example $\gamma$-rays, neutrinos and charged cosmic rays. These experiments have, thus far, not (definitively) detected DM particle candidates.
\\[10pt]
Dwarf spheroidal galaxies are promising targets for DM searches due to their high mass to light ratios (indicating an abundance of dark matter). Indirect dark matter searches for products of dark matter annihilation in dwarf spheroidal galaxies and the Galactic Centre have been conducted with a variety of observational facilities targeting different areas of particle parameter space. Claims of detection of a $\gamma$-ray spectral line signature \citep{Bringmann_and_Weinger_2012} and potential continuum emission \citep{Goodenough_and_Hooper_2009,Gordon_and_Macias_2013, Abramowski_et_al_2014} due to Weakly Interacting Massive Particles (WIMP) dark matter annihilation from dwarf spheroidal galaxies have been made. However, the fact that such claimed 
$\gamma$-ray signatures
have been, at best,
similar in magnitude to astrophysical $\gamma$-ray backgrounds -- 
which are
themselves somewhat uncertain -- has so far precluded conclusive identification of  observed $\gamma$-ray fluxes as products of dark matter annihilation \citep{Abramowski_et_al_2014,Geringer_Sameth_et_al_2015,Calore_et_al_2015b,Geringer_Sameth_et_al_2018, Macias_et_al_2018,Macias_et_al_2019,Abazajian_et_al_2020,Pohl:2022nnd}.
\\[10pt]
Compared to the Galactic Centre region and `classical' dwarf spheroidal (dSph) galaxies, which have been investigated extensively for indirect signatures of dark matter annihilation, few prior studies have previously investigated the Sagittarius Dwarf Spheroidal Galaxy \citep[hereafter Sgr;]{Ibata_et_al_1997, HESS:2007ora} as a target for dark matter annihilation searches. This is due to the location of Sgr near the Galactic plane and Galactic Centre region\footnote{$(l_{\rm Sgr},b_{\rm  Sgr})\approx(6^\circ,-14^\circ)$; \citet{Majewski_et_al_2003}.}, uncertain astrophysical background sources \citep{Viana_et_al_2012,Crocker_and_Macias_et_al_2022}, and large systematic uncertainties in the dark matter distribution (and thus the spatial morphology of any dark matter annihilation signature) of Sgr due to ongoing tidal disruption \citep{Rico_et_al_2020}. The most significant continuum detection of $\gamma$-ray emission from Sgr was made by \citet{Crocker_and_Macias_et_al_2022}, who detect Sgr with a $8.1\ \sigma$ significance 
in Fermi Large Area Telescope (LAT; \citealt{Atwood_et_al_2009}) data using their analysis pipeline. \citet{Crocker_and_Macias_et_al_2022} find the emitted Sgr $\gamma$-ray distribution spatially traces the stellar distribution of Sgr. The spectral distribution of Sgr $\gamma$-ray photons detected by \citet{Crocker_and_Macias_et_al_2022} strongly favour millisecond pulsar (hereafter MSP) $\gamma$-ray emission due to a combination of inverse Compton scattering of CMB photons by high-energy electron-positron pairs escaping from the Sgr MSP population and magnetospheric MSP $\gamma$-ray emission. Additionally, the H.E.S.S collaboration 
(marginally)
detected Sgr with a $2.05\ \sigma$ significance \citep{Abramowski_et_al_2014}; however, they conclude that the results are `well compatible' with a Gaussian significance distribution centred on zero. \citet{Viana_et_al_2012} also analyse Sgr for sources of potential $\gamma$-ray emission detectable with Cherenkov telescopes, concluding that predicted $\gamma$-ray emission from millisecond pulsars outshines the prospective $\gamma$-ray signal due to dark matter annihilation by several orders of magnitude.
\\[10pt]
This study seeks to predict the spatial and quantitative properties of the J-factor distribution of Sgr, providing a template for further searches for annihilation products from diverse particle physics models. In contrast to past studies utilising stellar tracers to estimate the Sgr dark matter density distribution and J-factor \citet{Viana_et_al_2012,Evans_et_al_2023}, we derive the J-factor distribution for Sgr from the hydrodynamic simulation of \citet{Tepper_Garcia_and_Bland_Hawthorn_2018}. This allows more accurate modelling of the considerable changes to the Sgr internal dynamics, DM density distribution and stellar density profile during the satellite's infall and tidal disruption \citep[Ch. 2]{Kazantzidis_et_al_2011,Newberg_and_Carlin_2016}. Section~\ref{sec:Simulation_overview} provides more information about these simulations, whilst the preliminary translation applied to the simulated particle distributions is detailed in section~\ref{sec:Particle_translation_method}. This translation does not change the resulting J-factor distribution magnitude. In section~\ref{sec:J_factor_calculation_method}, we adapt the methodology of \citet{Stoehr_et_al_2003} and \citet{Charbonnier_et_al_2011}, utilising the simulated Sgr dark matter particle mass and density values to produce a spatial J-factor flux distribution for use in indirect dark matter searches. The estimated volume-integrated J-factor value for Sgr is presented in section~\ref{sec:Predicted_J_factor_magnitude_result}, with further details of the spatial J-factor distribution of Sgr presented in sections~\ref{sec:1D_J_factor_profiles_results}. We then explore the implications of the J-factor value we derive in Section~\ref{sec:Discussion}, before concluding in section~\ref{sec:Summary_and_conclusions}.
\\[10pt]
To further inform observational searches, we also produce profiles of the simulated stellar density and dark matter density squared in section \ref{sec:Rho_profiles_section}. The former distribution provides an indication of the expected $\gamma$-ray emission from stellar-associated sources (for example millisecond pulsars) whilst the latter illustrates the DM density profile and the absolute magnitude of the Sgr J-factor distribution.
\section{Methodology}
\subsection{Simulation overview}
\label{sec:Simulation_overview}
Numerous simulations of the infall and tidal disruption of the progenitor galaxy of Sgr have been performed \citep[e.g.][]{Law_and_Majewski_2010,Lokas_et_al_2010,Law_et_al_2004,Dierickx_and_Loeb_2017}, differing in the initial position, mass and velocity of the Sgr progenitor and differing in the distribution of stellar and dark matter components. These difference in initial parameters has been shown to produce marked variations in the inferred orbit and evolution of the Sgr remnant \citep{Jiang_and_Binney_2000,Law_and_Majewski_2010,Lokas_et_al_2010}.
\\[10pt]
In contrast to prior simulations, the simulation of \citet{Tepper_Garcia_and_Bland_Hawthorn_2018} included a comprehensive treatment of gas in the Sgr progenitor. Realistic treatment of this gaseous component was shown to have a considerable effect on the orbital decay of Sgr, and also successfully reproduced other key features of the Sagittarius Dwarf/Stream system such as the approximate final position of stellar particles and the approximate angular size of the final Sagittarius Dwarf. Furthermore, in contrast to most previous simulations, this simulation includes initial conditions placing the Sgr progenitor at the virial radius of the Milky Way, which facilitates a more accurate treatment of the tidal disruption process during infall \citep{Dierickx_and_Loeb_2017}. This produced a more realistic evolution, including tidal stripping, of the Sgr Progenitor dark matter and stellar particles than simulations where the dwarf is artificially stripped and placed within the virial radius of the Milky Way. 
\\[10pt]
Here, we exploit this simulation to produce an estimation of the expected dark matter distribution of Sgr. As detailed in \citet{Tepper_Garcia_and_Bland_Hawthorn_2018}, this simulation included a Sagittarius Dwarf progenitor of total mass $11\times 10^{9}$ M$_\odot$ modelled with three live components. These components were a collisionless dark matter sub-halo of total mass $M_{DM} = 10^{10}$ M$_\odot$ (with a mass per particle of $10^5$ M$_\odot$), a collisionless stellar bulge of mass $M_{S} = 4\times 10^8$ M$_\odot$ (with a mass per particle of $4\times 10^3$ M$_\odot$) and a gaseous halo of mass $M_G = 6\times 10^8$ M$_\odot$ (and a mass per particle of $6\times 10^3$ M$_\odot$). Each of these components consisted of $10^5$ particles, with the initial mass distribution of these three components governed by a spherical Hernquist profile \citep{Hernquist_1990}. The Milky Way was also modelled as a live system, of total mass $1.087\times 10^{12}$ M$_\odot$, with five components (collisionless DM halo, collisionless stellar disk, collisionless stellar bulge, gas corona and gas disk) as detailed in Table 1 of \citet{Tepper_Garcia_and_Bland_Hawthorn_2018}.
\\[10pt]
The simulation was run utilising the version 3.0 of the Adaptive Mesh Refinement (AMR) scheme \scriptsize RAMSES \normalsize \citep{Teyssier_2002}. The Sgr progenitor was placed at an initial location of $\vec{r}_0=(125,0,0)$ kpc relative to the initial centre of the simulated Milky Way halo and had an initial velocity of $\vec{v}_0 \sim ($-$10,0,70)$ km s$^{-1}$. These initial conditions were adopted from the simulation of \citet{Dierickx_and_Loeb_2017}, where they resulted in the closet match (in the present-day configuration) between six simulated Sgr phase-space coordinates and the observed properties of Sgr, whilst also resulting in agreement between the simulated and observed position of the Sagittarius Stream stellar debris. The infall of the Sgr progenitor was simulated in infall for a total duration of $3.6$ Gyr, undertaking three pericentric passages, with the simulated stellar particles showing strong agreement with the observed distribution of Sagittarius Stream debris \citet{Tepper_Garcia_and_Bland_Hawthorn_2018}. For further details of the simulation, including details of the AMR implementation, sub-grid physics and a detailed description of the orbital evolution of the Sgr progenitor, see \citet{Tepper_Garcia_and_Bland_Hawthorn_2018}. For an extensive justification of the adopted initial conditions of the Sgr progenitor, see \citet{Dierickx_and_Loeb_2017}.
\subsection{Translation of the simulated Sagittarius Stream}
\label{sec:Particle_translation_method}
\begin{figure}
    \centering
    \includegraphics[width=\columnwidth]{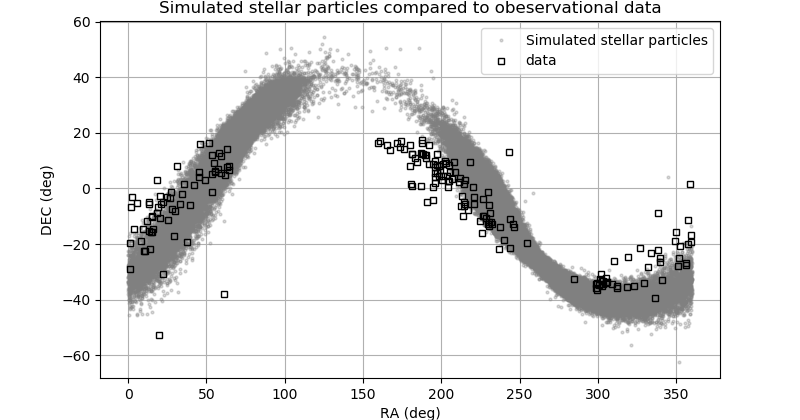}
    \caption{The simulated stellar particle population and observed stars \citep{Majewski_et_al_2003,Law_et_al_2005} in the Sagittarius Stream, reproduced from Figure 2 of \citet{Tepper_Garcia_and_Bland_Hawthorn_2018}.}
    \label{fig:F1_}
\end{figure}

\begin{figure}
    \includegraphics[width=\columnwidth]{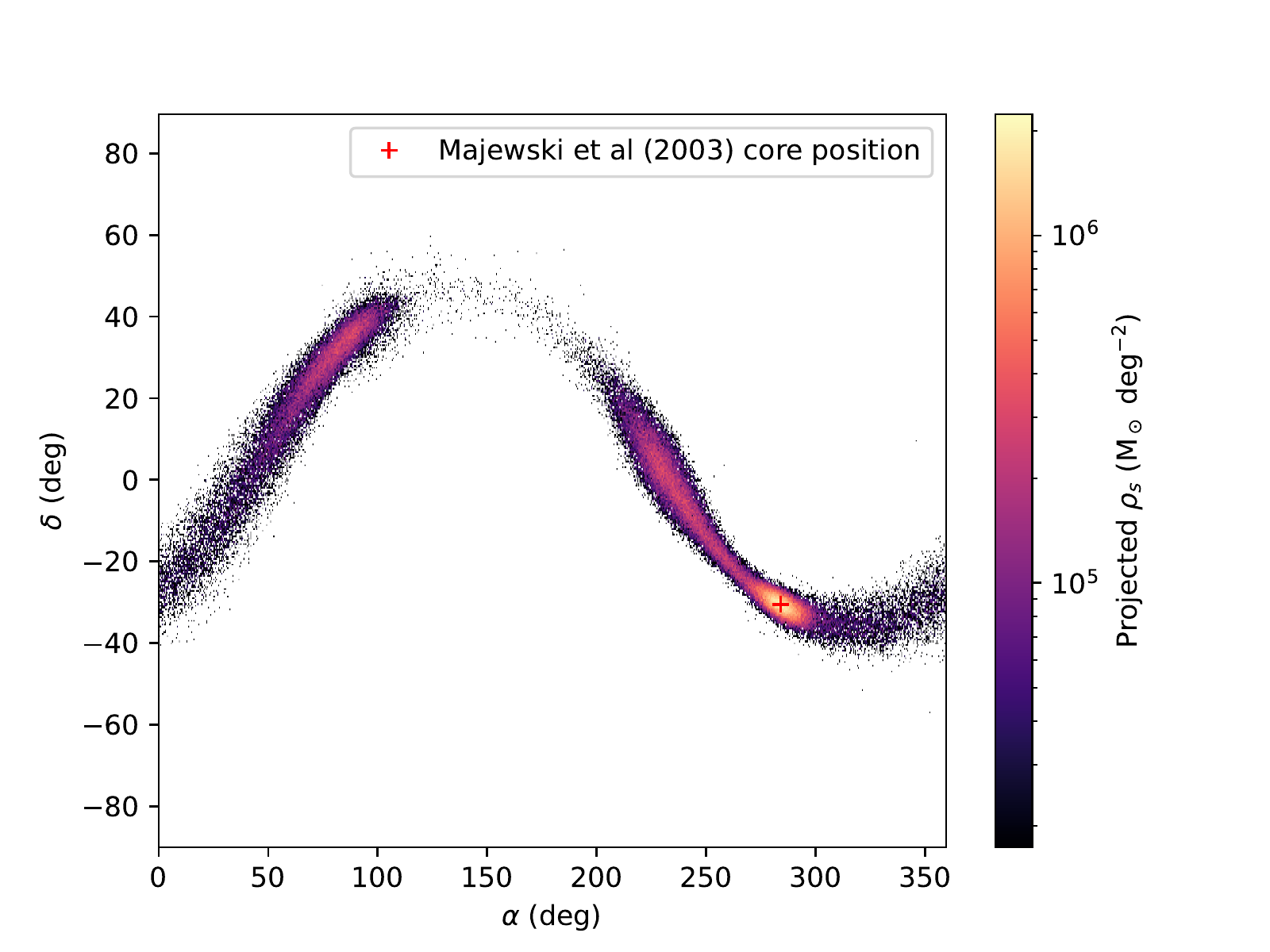}
    \caption{The translated projected mass density distribution of Sagittarius Stream simulated stellar particles. The position of these particles have been uniformly translated by $(\Delta \alpha,\Delta \delta) = (1.23,4.73)^\circ$ from their location in \citet{Tepper_Garcia_and_Bland_Hawthorn_2018}, and the projected location of the simulated Sgr is now $(\alpha,\delta) \simeq (284,-30.5)$. This is the location of Sgr reported by \citet{Majewski_et_al_2003}, as indicated by the red cross. As detailed in section~\ref{sec:Particle_translation_method}, this does not change the value of any projected quantities derived from the simulated particles.}
    \label{fig:F4}
\end{figure}

However, despite the successes of the \citet{Tepper_Garcia_and_Bland_Hawthorn_2018} model, it, in common with all previous simulations, fails to exactly reproduce the observed distribution of the Sagittarius Dwarf stars or the observed position of the Sgr Dwarf in the present day \citep{Majewski_et_al_2003,Belokurov_et_al_2014}, cf.~Figure~\ref{fig:F1_}. The projected location of the simulated Sgr (at the present day), defined as the position of the greatest projected stellar number (and mass) density\footnote{The location of Sgr reported in \citet{Tepper_Garcia_and_Bland_Hawthorn_2018}, $(\alpha,\delta) = (285,-36.6)^\circ$, is defined utilising the position of maximum 3D mass density.} occurred at approximately $(\alpha,\delta) \simeq (282.77,-35.23)^\circ$. This differs slightly from the observed location of Sgr reported in \citet{Majewski_et_al_2003}, $(l_{\rm Sgr},b_{\rm  Sgr})\approx(6^\circ,-14^\circ)$ or $(\alpha_{\rm Sgr},\delta_{\rm Sgr}) \simeq (284,-30.5)^\circ$.
\\[10pt]
An accurate density distribution of dark matter in the area of Sgr, informed by simulations, is of crucial importance to informing indirect dark matter searches. For spatial likelihood searches the position of Sgr is also crucial. Accordingly, to produce a dark matter and stellar template concordant with the observed Sagittarius Dwarf position, the simulated stars and dark matter particles from Sgr in the simulation of \citet{Tepper_Garcia_and_Bland_Hawthorn_2018} were translated in RA and DEC such that the position of maximum simulated stellar projected number (and mass) density of Sgr (in the centre of the simulated Sgr) was located at the observed location of Sgr detailed in \citet{Majewski_et_al_2003}. Specifically, this translated the simulated stellar and dark matter particle distributions by $(\Delta \alpha,\Delta \delta) = (1.23,4.73)^\circ$; the location of the simulated Sgr was moved from a projected location of $(\alpha,\delta) \simeq (282.77,-35.23)^\circ$ to $(\alpha',\delta') \simeq (284,-30.5)^\circ$. 
The heliocentric distance of Sgr remained unchanged at $25.3$ kpc following this translation process. Note that this translation preserves the 3D structure of the simulated stream, given this translation preserves the relative position of all Sgr particles and does not result in a change in heliocentric distance to Sgr. Accordingly, any quantities calculated from the simulated particle distributions remain unchanged in magnitude as a result of this transformation.
Figure~\ref{fig:F4} illustrates the translated, projected simulated stellar mass density distribution. The projected mass density for each pixel was defined as the sum of particle masses divided by the angular area of the pixel:
\begin{equation}
    \rho_s = \frac{1}{\alpha^2} \sum_i m_i
    \label{eq:Simulated_Stellar_projected_mass_density}
\end{equation}
where $m_i$ is the mass of particle $i$ and $\alpha$ is the angular width of the pixels. A constant square pixel side length of $\alpha=0.458^\circ$ was adopted for when calculating all projected distributions, which corresponds to a physical size of $0.0035$ kpc at the distance of the Sagittarius Dwarf ($25.3$ kpc). This corresponds to a square pixel size of $\alpha^2=0.21$ square degrees. This square pixel size was adopted to facilitate use of these distributions in a maximum-likelihood analysis of Fermi-LAT data as the angular resolution of the Fermi-LAT instrument at $1$ GeV is approximately equal to this adopted pixel size.  
\\[10pt]
We used the translated projected dark matter particle distribution to calculate the Sgr J-factor distribution from the resulting projected Sagittarius Dwarf density profile. These calculations (detailed in section~\ref{sec:J_factor_calculation_method}) followed the method of \citet{Stoehr_et_al_2003} to calculate the relevant distribution directly from simulated particle properties (also see \citealt{Kuhlen_2009,Charbonnier_et_al_2011}).
\subsection{Production of projected J-factor distributions}
\label{sec:J_factor_calculation_method}
\begin{figure}
    \includegraphics[width=\columnwidth]{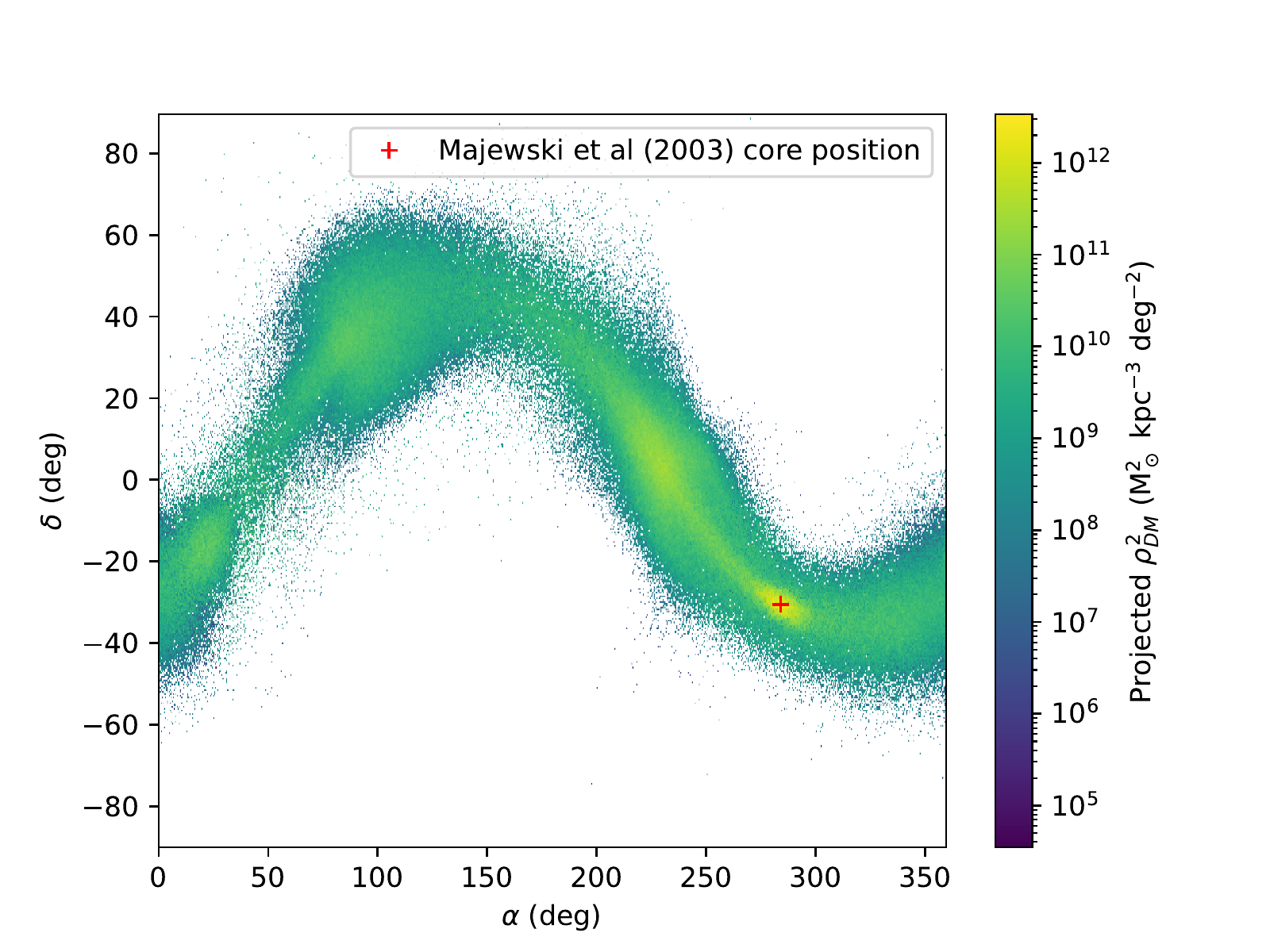}
    \caption{The projected $\rho_{DM}^2$ distribution of the translated Sagittarius Stream. As per the simulated stellar translation, all particles are translated by $(\Delta \alpha,\Delta \delta) = (1.23,4.73)^\circ$ from their detailed location in \citet{Tepper_Garcia_and_Bland_Hawthorn_2018}. As detailed in section~\ref{sec:Particle_translation_method}, this does not change the value of any projected quantities derived from the simulated particles. The observed location of Sgr reported in \citet{Majewski_et_al_2003} is indicated with a red cross, which is coincident with the translated simulated Sgr dark matter population.}
    \label{fig:F5}
\end{figure}

\begin{figure}
    \centering 
    \includegraphics[width=\columnwidth]{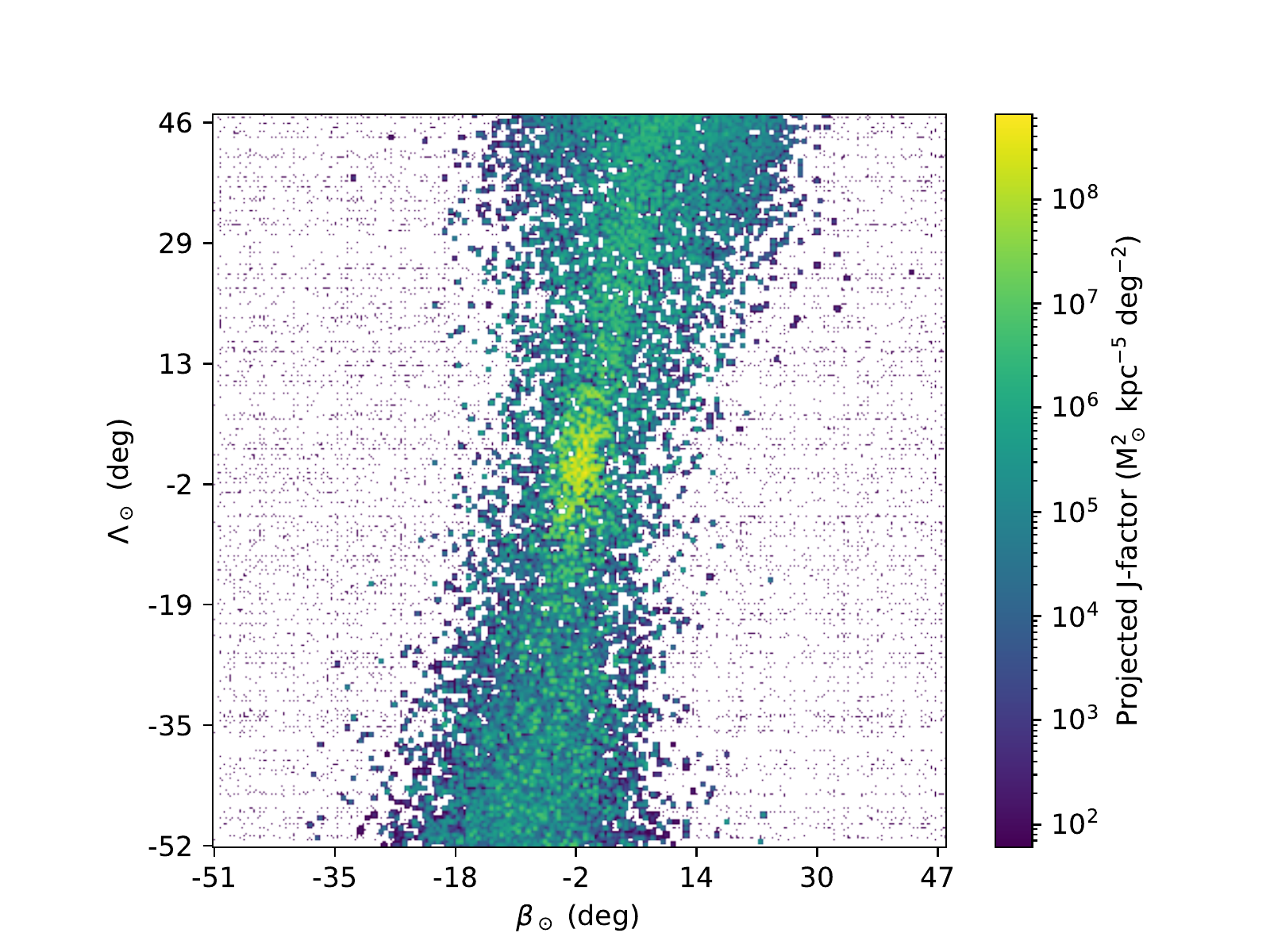}
    \caption{The projected J-factor distribution centred on the location of the simulated Sagittarius Dwarf, depicted in the coordinate system of \citet{Belokurov_et_al_2014}. The pixel size adopted in this figure is $\alpha^2=0.21$ square degrees. The location of Sgr in this coordinate system is $(\beta_{\text{Sgr}},\Lambda_{\text{Sgr}})= (-1.48,-0.22)^\circ$.}
    \label{fig:J_factor_DM_ROI_angular_units}
\end{figure}

The self-annihilation or decay of several families of particle dark matter candidates is generically expected to produce $\gamma$-ray emission \citep[e.g.,][]{Bertone_2005}. In the case of WIMPs \citep[][]{Jungman_1996}, primary and secondary $\gamma$-rays are produced as products of self annihilation through multiple production channels (in addition to other annihilation products), with the volumetric annihilation rate scaling with the square of mass density. For astrophysical targets such as dwarf spheroidal galaxies and the Galactic centre the expected $\gamma$-ray flux from WIMP self-annihilation, as a function of energy $E$ per unit energy per solid angle $\Omega$, can be modelled using an equation of the form \citep{Charbonnier_et_al_2011}:
\begin{equation}
    \frac{d\Phi_{\gamma}}{dE_\gamma}(E,\Delta \Omega)= \frac{1}{4\pi}\frac{\langle\sigma\nu\rangle}{2m^2}\frac{dN_\gamma}{dE_\gamma}(E)\times J(\Delta \Omega)
    \label{EQ1}
\end{equation}
where the velocity averaged DM annihilation cross section $\langle\sigma\nu\rangle$, DM particle mass $m$ and spectral energy distribution of emitted $\gamma$-rays $\frac{dN_\gamma}{dE}(E)$ are model dependent parameters. Together, these terms detail the spectral distribution of the $\gamma$-ray annihilation products.
In contrast, the `J-factor' $J(\Delta \Omega)$ specifies the spatial dependence of the $\gamma$-ray flux. Explicitly in the case of WIMP annihilation \citep{Charbonnier_et_al_2011},
\begin{equation}
    J = \int_{\Delta \Omega}\frac{dJ}{d\Omega} = \int_{\Delta \Omega}\int \rho^2(r,\Omega) \text{d}r\text{d}\Omega
    \label{EQ2}
\end{equation}
where $\rho$ is the density distribution of dark matter along the line of sight radii $r$ to the object of angular size $\Delta \Omega$.
\\[10pt]
However, accurately constraining the J-factor value for the dark matter component of Sgr is difficult due to ongoing tidal disruption (and uncertain foreground/background $\gamma$-ray sources). This results in a dark matter density distribution subject to significant uncertainties, comparatively difficult (compared to classical dSph systems) to constrain with stellar tracers. Simulations are often used to model the dark matter density distribution of Sgr; however, to the knowledge of the authors no simulation has been utilised to model the J-factor distribution and compute an integrated J-factor value for Sgr.
\\[10pt]
The quantity $\rho_{DM}^2$, the dark matter spatial density squared integrated over a given volume, was estimated directly from the simulation output particle density $\rho_i$ in accordance with the equation \citep{Stoehr_et_al_2003}:
\begin{equation}
    \rho_{DM}^2 = \int_{V} \rho_{DM}^2\ d\text{V} = \sum_i \rho_i m_i
\label{EQ8}
\end{equation}
where $m_i$ is the (constant) simulated particle mass. The projected distribution of this quantity for the Sagittarius Stream is shown in Figure~\ref{fig:F5}, calculated for each pixel by summing over all particles in the pixel in accordance with equation~\ref{EQ8} before division by the pixel size ($\alpha^2=0.21$ square degrees).
Note that whilst \citet{Stoehr_et_al_2003} utilise equation~\ref{EQ8} to calculate the J-factor value for a given volume element, in this study J-factor values for substructures within the simulation were calculated including an inverse-square distance dependence given the extended nature of Sgr. Specifically, we include a factor of $1/(4\pi r_i^2)$ for each particle $i$ to account for the size of the flux sphere for each simulation particle. With this distance dependence, the J-factor definition adopted in this study is equivalent to a flux, not absolute brightness. The contribution of each particle $i$ to the J-factor value for each volume element $V_b$ at a heliocentric distance $r$ of the considered structure was thus calculated as:
\begin{equation}
    J_b = \int_{V_b} \rho_{DM}^2/(4\pi r^2)\ d\text{V} = \sum_i \rho_i m_i/(4\pi r_i^2)
\label{eq:EQ6}
\end{equation}
where $\rho_i$ is the dark matter density, $m_i$ is the mass of particle $i$ and $r_i$ is the distance from the Sun to particle $i$. As shown in appendix~\ref{sec:Charbonnier_equivalence_appendix}, this definition is equivalent to the J-factor definition provided by \citet{Charbonnier_et_al_2011}, though differs by a factor of $4\pi$ due to the flux sphere surface area normalization factor. The total summed J-factor value for an extended substructure (for example Sgr) was calculated by summing over the constituent volume elements $V_b$: 
\begin{multline}
    J = \Sigma_b \int_{V_b} \rho_{DM}^2/(4\pi r^2)\ d\text{V} = \Sigma_b\Sigma_i \rho_i m_i/(4\pi r_i^2) \\ = \Sigma_{i'}\rho_{i'} m_{i'}/(4\pi r_{i'}^2) 
\label{EQ9}
\end{multline}
for all particles $i'$ in the extended substructure (potentially of multiple volume elements).
\\[10pt]
For the analysis of this study, projected distributions of the line of sight integrated J-factor values were produced using the HEALPix pixelization scheme \citep{Gorski_et_al_2005} to sum the particle values in spatial pixels (volume elements). The integrated J-factor value for each pixel was calculated using equation~\ref{eq:EQ6} and converted into a projected value through division by the pixel size. The side width of the square bins adopted in this plot, and in all 2D plots in this paper, was $\alpha=0.458^\circ$, which corresponds to a square pixel area of $\alpha^2=0.21$ square degrees. As aforementioned, this square pixel size corresponds to the resolution of the Fermi-LAT instrument at $1$ GeV and was chosen to allow the use of these projected distributions in analyses of Fermi-LAT data. Figure~\ref{fig:J_factor_DM_ROI_angular_units} illustrates the projected dark matter J-factor distribution at the location of the Sagittarius core in the units of M$_\odot^2$ kpc$^{-5}$ deg$^{-2}$.
In Figure~\ref{fig:J_factor_DM_ROI_physical_units}, this distribution is also shown in the physical units of M$_\odot^2$ kpc$^{-7}$, converted from the units of M$_\odot^2$ kpc$^{-5}$ deg$^{-2}$ through division of each pixel $b$ by the factor $C_b$:
\begin{equation}
    C_b = (2\tan(\alpha/2)\times \Bar{d}_b)^2/\alpha^2
    \label{eq:C}
\end{equation}
where $\alpha=0.458^\circ$ was the constant angular width of each pixel and $\Bar{d}_b$ was the mean particle distance in the given pixel $b$. We have verified that the standard deviation of $C_b$ for each pixel remains significantly below the value of $C_b$, and accordingly that the transformation is robust to the variance of particle distances about the mean. In these figures, we use the ($\beta_\odot,\Lambda_\odot$) coordinate system of \citet{Belokurov_et_al_2014}, which is a variant of the coordinate system utilised in \citet{Majewski_et_al_2003}. In the system of \citet{Belokurov_et_al_2014}, the $\Lambda_\odot$ axis proceeds along the Sagittarius Stream in the direction of motion of Sgr, whilst the $\beta_\odot$ axis points toward the North Galactic Pole. These figures detail the region $\Lambda_\odot \in [\Lambda_{0}\pm 50^\circ], \beta_\odot \in [\beta_{0} \pm 50\circ]$, where $(\beta_{0},\Lambda_{0}) = (-1.48,-2.22)$. Sgr is located at $(\beta_{\text{Sgr}},\Lambda_{\text{Sgr}})= (-1.48,-0.22)^\circ$ in this coordinate system.
\\[10pt]
In addition to this paper analyzing the J-factor distribution of Sgr, an upcoming study (Venville, in prep, hereafter Paper II) will discuss the all-sky J-factor distribution of the Sagittarius Stream and investigate the possibility of detecting $\gamma$-ray emission from the Sagittarius Stream utilising Fermi LAT data.
\section{Results: predicted dark matter density and J-factor distributions from the Sagittarius Dwarf}
\subsection{Predicted J-factor magnitude of the Sagittarius Dwarf}
\label{sec:Predicted_J_factor_magnitude_result}
In accordance with equation~\ref{EQ9}, the integrated J-factor value for Sgr was calculated by summing all particle values along the line of sight and within the projected `core radius' (corresponding to the approximate bound radius) of $\alpha_{\text{Sgr}} \simeq 3.7^\circ$ \citep{Majewski_et_al_2003} of the centre of the simulated Sagittarius Dwarf, defined as $(\alpha,\delta) \simeq (284,-30.5)^\circ$ in section~\ref{sec:Particle_translation_method} and \citet{Majewski_et_al_2003}. This radius corresponds to a physical size of $1.6$ kpc at the location of Sgr and a conical integration region of solid angle of $\Omega_{\text{Sgr}} \simeq 3.6\times 10^{-3}$ sr.
\\[10pt]
This integrated J-factor value of Sgr was $1.48\times 10^{10}$ M$_\odot^2$ kpc$^{-5}$. Note that this is not a projected J-factor value but (in accordance with equation~\ref{EQ9}) instead corresponds to a three dimensional volume summation of the particle J-factor values within this conical integration region. It should also be noted that due to the archival nature of the utilised simulation, the convergence of the derived Sgr particle density profile at varying resolutions cannot be studied. As increasing resolution \textit{could} result in an increasing Sgr central density and correspondingly increased J-factor values at small radii, the J-factor value derived in this study is a conservative lower limit.
\\[10pt]
Prior determinations of the J-factor value for Sgr are consistently higher than value derived in this paper. \citet{Abramowski_et_al_2014} estimate a J-factor value of $J_{\text{Sgr}} = 2.88\times 10^{12}$ M$_\odot^2$ kpc$^{-5}$ integrated across a solid angle of $\Delta \Omega = 10^{-5}\ \text{sr} \simeq 0.032\ \text{deg}^2$, whereas \citet{Viana_et_al_2012} calculate a J-factor value of $J_{\text{Sgr}} = 1.5\times 10^{15}$ M$_\odot^2$ kpc$^{-5}$ for an integration solid angle of $\Delta \Omega = 2\times10^{-5}\ \text{sr} = 0.07\ \text{deg}^2$. These results shall be discussed in Section~\ref{sec:Discussion}.
\\[10pt]
The projected J-factor distribution for Sgr produced in this study was fitted (in conjunction with the additional templates described in \citet{Crocker_and_Macias_et_al_2022}) to Fermi-LAT $\gamma$-ray data utilising the exact methodology detailed in \citet{Crocker_and_Macias_et_al_2022}). The normalized DM J-factor distribution was detected with a maximum
significance of $<1\sigma$, including during evaluation of transitional and rotational tests. This DM template is thus significantly less favoured by the data than templates tracing the Sgr projected stellar density distribution (detected with an $8.1\ \sigma$ significance). Therefore, \citet{Crocker_and_Macias_et_al_2022} concluding that MSPs are the likely cause of $\gamma$-ray emission from the direction of Sgr is consistent both with the findings of \citet{Viana_et_al_2012} and the predicted J-factor distribution for Sgr calculated in this study. To provide further support for this conclusion, we utilise the J-factor value calculated here and the $\gamma$-ray photon flux attributed to Sgr in \citet{Crocker_and_Macias_et_al_2022} to calculate a lower limit on the required WIMP cross-section $\langle \sigma \nu \rangle$ to produce this $\gamma$-ray photon flux. In accordance with equation~\ref{EQ1} and \citet{Mazziotta_et_al_2012}, the photon flux $\Phi_\gamma$ can be expressed as
\begin{equation}
    \Phi_\gamma(E,\Delta\Omega) = J(\Delta\Omega)\times\frac{1}{2}\frac{\langle\sigma\nu\rangle}{4\pi m^2_\chi}\Sigma_f N_f(E,m_\chi)B_f
\end{equation}
where $m_\chi$ is the mass of the WIMP and $N_f(E,m_\chi)$ is the differential photon spectrum produced by pair annihilation into a final state $f$ with a branching fraction $B_f$. Setting $N_f=B_f=1$ allows the calculation of a lower limit on the cross section value for an assumed WIMP mass and J-factor value $J(\Delta\Omega)$. For 
the J-factor of magnitude $J_{\text{Sgr}} \sim 10^{10}$ M$_\odot^2$ kpc$^{-5}$ as determined in this study, the lower limit on the cross section required to produce the GeV-band $\gamma$-ray number flux $\Phi_\gamma
\sim 10^{-8}\ \text{cm}^{-2}\ \text{s}^{-1}$ observed in \citet{Crocker_and_Macias_et_al_2022} is $\langle \sigma \nu \rangle \sim 6 \times10^{-20} \text{cm}^{3}\ \text{s}^{-1} \left(\frac{m_\chi}{\rm 100 \ GeV}\right)^2$. Such a velocity averaged cross-section is inconsistent  with existing constraints on WIMP cross sections (\citealt[Figure 1]{Abazajian_et_al_2020}; \citealt[Figure 10, left]{Albert_et_al_2017}). For example, assuming a WIMP mass of $10$ GeV, the lower limit on the required cross-section implied by our study is $\langle \sigma \nu \rangle \sim 6\times10^{-24}\ \text{cm}^{3}\ \text{s}^{-1}$, far in excess of the upper limit of $\langle \sigma \nu \rangle \sim 3 \times10^{-27} \text{cm}^{3}\ \text{s}^{-1}$ derived from analysis of other measured dSph galaxy J-factors \citep[Figure 10, left]{Albert_et_al_2017}. It also significantly exceeds the upper limit of $\langle \sigma \nu \rangle \sim 3 \times10^{-28} \text{cm}^{3}\ \text{s}^{-1}$ derived from analysis of the Galactic Centre Excess (assuming a NFW DM density distribution) detailed in \citet[Figure 1]{Abazajian_et_al_2020}. Similarly, for a WIMP mass of $100$ GeV, the lower limit of $\langle \sigma \nu \rangle \sim 6\times10^{-20}\ \text{cm}^{3}\ \text{s}^{-1}$ implied by this study far exceeds the upper limits of $\langle \sigma \nu \rangle \sim 1.5\times10^{-26}\ \text{cm}^{3}\ \text{s}^{-1}$, derived from analysis of other measured dSph galaxy J-factors in \citet{Albert_et_al_2017}, and $\langle \sigma \nu \rangle \sim 2\times10^{-27}\ \text{cm}^{3}\ \text{s}^{-1}$, derived from analysis of the Galactic Centre Excess in \citet{Abazajian_et_al_2020}. This incompatibility of the implied WIMP DM velocity averaged cross-section found by this study with existing constraints reinforces the conclusions of \citet{Crocker_and_Macias_et_al_2022} and strongly disfavours WIMP DM as the source of the observed $\gamma$-ray emission from the Sgr.
\subsection{1D J-factor profiles}
\label{sec:1D_J_factor_profiles_results}
As the J-factor value for Sgr in this study is low compared to prior estimations \citep{Abramowski_et_al_2014,Viana_et_al_2012} and significant $\gamma$-ray emission from MSPs is present at the location of Sgr \citep{Crocker_and_Macias_et_al_2022}, morphological features of the Sgr J-factor distribution may aid detection of any $\gamma$-ray emission due to DM annihilation. Thus, the following section shall detail morphological characteristics of the DM J-factor distribution inferred from the simulations. This has the potential advantage, compared to DM distributions derived from stellar distribution data \citep[e.g.][]{Viana_et_al_2012}, of more accurate treatment of the tidal disruption of the DM component and avoiding the need to assume dynamical equilibrium or spherical symmetry in the DM (or stellar tracer) population.
\begin{figure*}
    \centering
    \includegraphics[width=2\columnwidth]{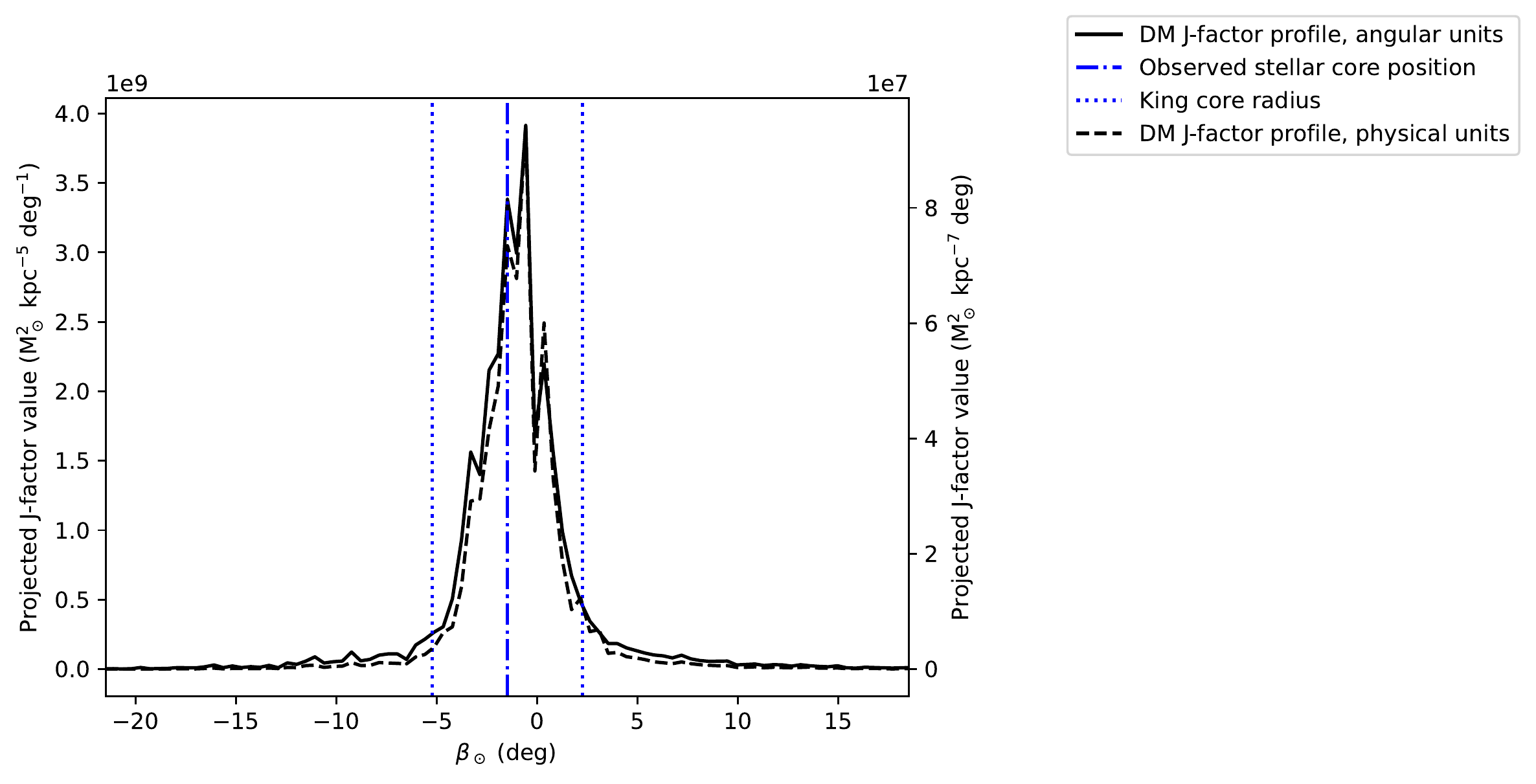}
    \caption{The projected J-factor profile for the dark matter particles as a function of $\beta_\odot$, within the region of interest defined in Section~\ref{sec:J_factor_calculation_method} and depicted in Figure~\ref{fig:J_factor_DM_ROI_angular_units}. The adopted $\beta_\odot$ bin size is $0.458^\circ$. This region of interest spans the range $-50^\circ<\Lambda_\odot<50^\circ$. Note that this J-factor profile is depicted in both the angular units of M$_\odot^2$ kpc$^{-5}$ deg$^{-2}$ (derived from the distribution depicted in Figure~\ref{fig:J_factor_DM_ROI_angular_units}) and the physical units of M$_\odot^2$ kpc$^{-7}$ (derived from the distribution depicted in Figure~\ref{fig:J_factor_DM_ROI_physical_units}).}
    \label{fig:DM_J_factor_Beta_profile}
\end{figure*}
\\[10pt]
The projected J-factor distribution summed in bins of $\beta_\odot$ over the region of interest detailed in section~\ref{sec:J_factor_calculation_method} is illustrated in Figure~\ref{fig:DM_J_factor_Beta_profile}. The bin width adopted for this binning, and all subsequent 1 dimensional profiles in this paper, was $0.458^\circ$. This bin width is equal to the angular resolution of the Fermi-LAT instrument at $1$ GeV. It is an intentionally identical width to the width/length of the square pixels utilised in calculating the projected J-factor distributions detailed in section~\ref{sec:J_factor_calculation_method}; as aforementioned, the square pixel size of these projected distributions was chosen to allow use of these distributions in a maximum-likelihood analysis of Fermi-LAT data. As is evident in Figure~\ref{fig:DM_J_factor_Beta_profile}, there is a marked increase in the magnitude of the predicted J-factor at the location of Sgr; however, $\gamma$-ray emission from stellar populations or millisecond pulsars are also expected to increase in magnitude at this location \citep{Viana_et_al_2012}.
\\[10pt]
The magnitude of the projected J-factor profile depicted as a function of $\beta_\odot$ in Figure~\ref{fig:DM_J_factor_Beta_profile} at the observed location of Sgr is $3.4\times10^9$ M$_\odot^2$ kpc$^{-5}$ deg$^{-1}$; this is approximately $13$ times the J-factor value of $2.6\times10^8$ M$_\odot^2$ kpc$^{-5}$ deg$^{-1}$ at the core radius of Sgr in the negative $\beta_\odot$ direction, namely ($\beta_\odot,\Lambda_\odot)=(-5.21,0)^\circ$. At the core radius of Sgr, but in the positive $\beta_\odot$ direction, the J-factor value is $4.6\times10^8$ M$_\odot^2$ kpc$^{-7}$ deg, a factor of approximately $7.3$ lower than the J-factor value at the observed location of Sgr. The difference between these values indicates the relative J-factor contrast between the core and outskirts of Sgr. This contrast is significantly lower than the stellar number density contrast across similar angular scales in the fitted templates adopted in \citet[Supplementary Figure 1]{Crocker_and_Macias_et_al_2022}. Accordingly, the DM J-factor templates are likely insufficiently peaked to fit the spatial $\gamma$-ray emission distribution observed for Sgr.
\\[10pt]
Similarly, the projected J-factor profile for the DM particles summed in bins of $\Lambda_\odot$ is illustrated in Figure~\ref{fig:DM_J_factor_Lambda_profile}. The value of this distribution at the observed core central position is $1.3\times10^9$ M$_\odot^2$ kpc$^{-5}$ deg$^{-1}$, whilst the J-factor value at the core radius in negative $\Lambda_\odot$ direction$(\beta_\odot,\Lambda_\odot)=(0,-4.0)^\circ$ is $6.9\times10^8$ M$_\odot^2$ kpc$^{-5}$ deg$^{-1}$. The J-factor value at the core radius in the positive $\Lambda_\odot$ direction, at $(\beta_\odot,\Lambda_\odot)=(0,3.51)^\circ$, is $9.5\times10^8$ M$_\odot^2$ kpc$^{-5}$ deg$^{-1}$. However, the $\Lambda_\odot$ profile detailed in Figure~\ref{fig:DM_J_factor_Lambda_profile} shows significant asymmetry and variance on small angular scales, resulting from the tidal disruption of the Sgr halo. Regardless, this J-factor profile is also likely insufficiently peaked to fit the spatial $\gamma$-ray distribution attributed to Sgr in \citet{Crocker_and_Macias_et_al_2022}.
\\[10pt]
\begin{figure*}
    \centering
    \includegraphics[width=2\columnwidth]{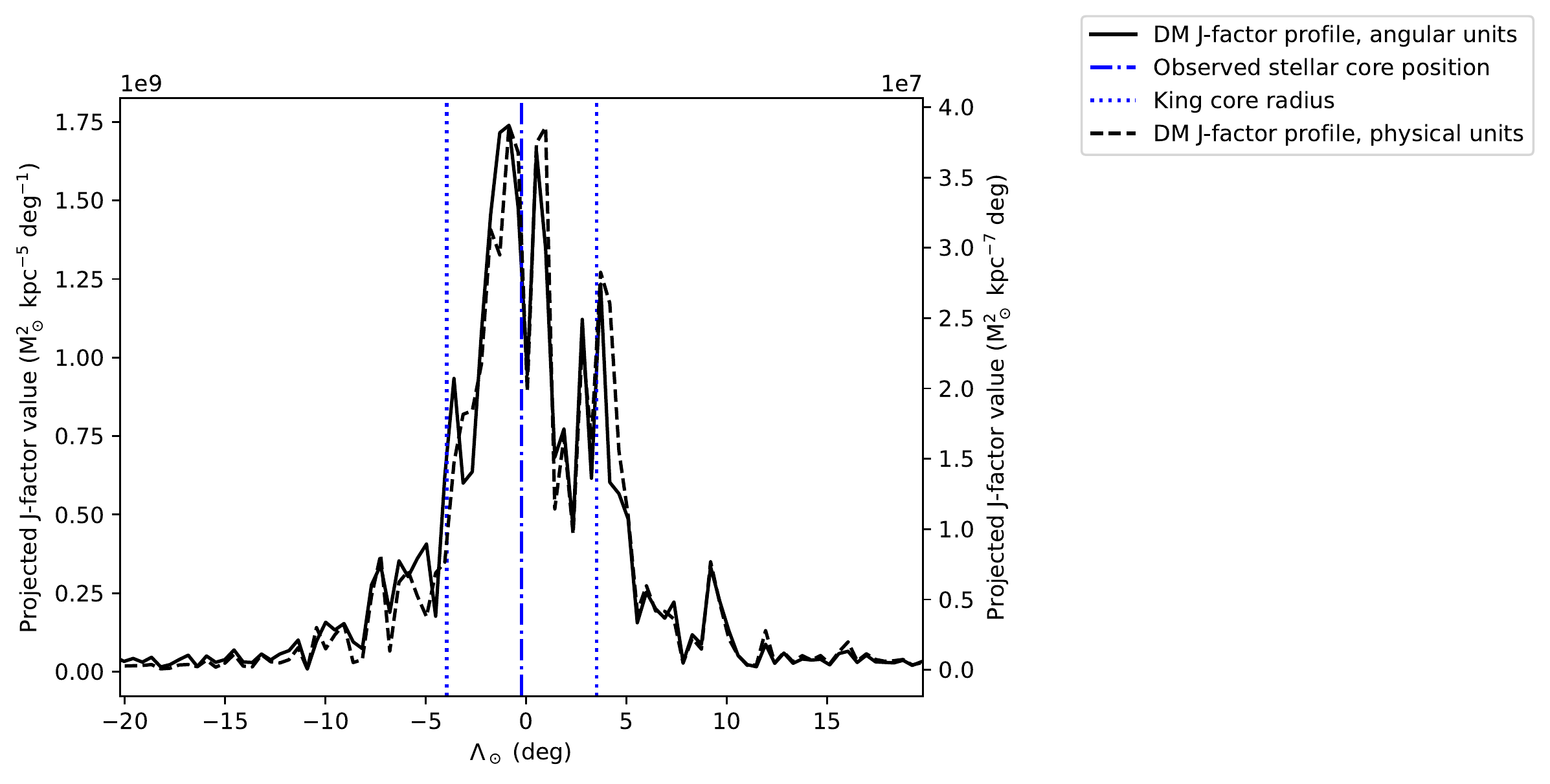}
    \caption{The projected J-factor profile for the dark matter particles as a function of $\Lambda_\odot$, within the region of interest defined in Section~\ref{sec:1D_J_factor_profiles_results}. This region of interest spans the range $-50^\circ<\beta_\odot<50^\circ$. The adopted $\Lambda_\odot$ bin size is $0.458^\circ$. As per the J-factor profile as a function of $\beta_\odot$ depicted in Figure~\ref{fig:DM_J_factor_Beta_profile}, the J-factor profile is depicted in both the angular units of M$_\odot^2$ kpc$^{-5}$ deg$^{-2}$ (derived from the distribution depicted in Figure~\ref{fig:J_factor_DM_ROI_angular_units}) and the physical units of M$_\odot^2$ kpc$^{-7}$ (derived from the distribution depicted in Figure~\ref{fig:J_factor_DM_ROI_physical_units}).}
    \label{fig:DM_J_factor_Lambda_profile}
\end{figure*}
To compare the width of the 1D projected J-factor profiles to observed stellar distributions and simulated DM/stellar density distributions (see section~\ref{sec:Rho_profiles_section}), distribution functions were fitted to scaled versions of the projected J-factor profiles with a maximum value of $1.0$, as illustrated in Figure~\ref{fig:DM_J_factor_Beta_profile_normalized} for the dark matter J-factor $\beta_\odot$ profile and Figure~\ref{fig:DM_J_factor_Lambda_profile_normalized} for the $\Lambda_\odot$ profile. We choose to scale the profiles with angular units to avoid any small scale variation caused on the profiles with physical units by variation in the applied divisor $C_b$ between pixels. The selected form of these functions, either a Voigt, Moffat, Gaussian or Gaussian with constant vertical offset, were selected by minimizing the following non-linear least squares optimization function value $O$:
\begin{equation}
    O=\sum_x \left(y'(x)-y(x)\right)^2
    \label{eq:Opt_function}
\end{equation}
where $y'$ is the value of the fitted function and $y$ is the value of the 1D profile at a given value of $x=\beta_\odot$ or $x=\Lambda_\odot$, depending on the 1D profile in question. These functions were fitted to the 1D J-factor profiles using a Levenberng-Marquardat \citep{Levenberg_1944,Marquardt_1963} non-linear least squares algorithm, as implemented in \scriptsize ASTROPY\footnote{The fitting function implemented in \scriptsize ASTROPY \footnotesize is detailed \href{https://docs.astropy.org/en/stable/api/astropy.modeling.fitting.LevMarLSQFitter.html}{at this website}.} \normalsize (\citealt{astropy:2013,astropy:2018,astropy:2022}, in the case of the Moffat, Gaussian or Voigt functions) or \scriptsize SCIPY\footnote{The fitting function implemented in \scriptsize SCIPY \footnotesize is detailed \href{https://docs.scipy.org/doc/scipy/reference/generated/scipy.optimize.curve\_fit.html}{at this website}.} \normalsize (\citealt{Virtanen_et_al_2020}, in the case of the offset Gaussian function). A Moffat function of the form 
\begin{equation}
    f(x) = A \left(1 + \frac{\left(x - x_{0}\right)^{2}}{\gamma^{2}}\right)^{-\alpha}
    \label{eq:Moffat}
\end{equation}
best fit both the scaled $\beta_\odot$ and scaled $\Lambda_\odot$ J-factor profiles as it more accurately models the wide tails of these distributions resulting from the tidal disruption of Sgr. The parameters of these fits are detailed in table~\ref{tab:Fits_summary_table}. The full width half maximum (FWHM) of the Moffat function fitted to the scaled J-factor profile along the $\beta_\odot$ axis is $3.7^\circ$, comparable to the observed `core radius' of Sgr reported in \citet{Majewski_et_al_2003}. The FWHM of the Moffat function fitted to the scaled DM $\Lambda_\odot$ J-factor profile is $4.9^\circ$. This illustrates the known tidal disruption of Sgr along the $\Lambda_\odot$ axis \citep{Majewski_et_al_2003,Lokas_et_al_2010}. The Sgr J-factor distribution is thus notably extended along both $\beta_\odot$ and $\Lambda_\odot$ axes, providing a valuable morphological discriminant to point-like sources of emission present in analysis procedures given any $\gamma$-ray emission from the Sgr DM population would follow a similarly extended distribution.
\begin{table*}
    \centering
    \caption{The parameters of the distribution functions fitted to scaled 1D projected J-factor, simulated stellar density ($\rho_s$) and DM density squared ($\rho_{DM}^2$) distributions. These functions were fitted to provide insight into the morphology of expected $\gamma$-ray emission from DM annihilation (in the case of the J-factor distributions) and stellar-associated sources (in the case of the $\rho_s$ distributions). The $\rho_{DM}^2$ distributions provide insight into the tidally disrupted Sgr DM halo. The scaled J-factor distributions and fitted functions are further discussed in section~\ref{sec:1D_J_factor_profiles_results}, whilst the scaled $\rho_s$ and $\rho_{DM}^2$ distributions and fitted functions are discussed in section~\ref{sec:Rho_profiles_section}.}
    \begin{tabular}{c|c|c|c|c}
        Profile & Fitted distribution & Fitted parameters & FWHM & Optimization function value \\
        \hline
        \hline
        DM J-factor - $\beta_\odot$ & Moffat & $A=0.854$ & $3.7^\circ$ & $0.141$ \\
         & &  $x_0 = -1.07$ & \\
         & & $\gamma=2.12$ & \\
         & & $\alpha=1.50$ & \\
        \hline
        DM J-factor - $\Lambda_\odot$ & Moffat & $A=0.885$ & $4.9^\circ$ & $0.745$ \\
         & &  $x_0 = -0.437$ & \\
         & & $\gamma=2.70$ & \\
         & & $\alpha=0.92$ & \\
        \hline
        $\rho_s$ - $\beta_\odot$ & Moffat & $A=0.941$ & $4.9^\circ$ & $0.181$ \\
         & &  $x_0 = -0.757$ & \\ 
         & & $\gamma=2.70$ & \\
         & & $\alpha=1.06$ & \\
        \hline
        $\rho_s$ - $\Lambda_\odot$ & offset Gaussian & $a=8.65$ & $13.6^\circ$ & $0.875$\\
         & &  $mu=-0.390$ & \\ 
         & & $\sigma=5.10$ & \\
         & & $c=0.094$ & \\
        \hline
        $\rho_{DM}^2$ - $\beta_\odot$ & Moffat & $A=0.873$ & $3.6^\circ$ & $0.145$ \\
         & &  $x_0 = -1.04$ & \\ 
         & & $\gamma=1.57$ & \\
         & & $\alpha=0.978$ & \\
        \hline
         $\rho_{DM}^2$ - $\Lambda_\odot$ & offset Gaussian & $a=5.62$ & $8.0^\circ$ & $0.838$\\
         & &  $mu=-0.196$ & \\ 
         & & $\sigma=3.07$ & \\
         & & $c=0.041$ & \\
        \hline
    \end{tabular}
    \label{tab:Fits_summary_table}
\end{table*}
\\[10pt]
We also conduct an estimation of the percentage contribution of any small scale profile irregularities in the 1D projected J-factor profiles to the overall J-factor value. To compute this contribution, we compute a J-factor value by integrating the fitted 1D distributions over a given interval, and compare this to the J-factor values found by numerically integrating the 1D projected profile bins over the same interval. This numerical integral utilised the cumulative trapezoid method, as implemented in the \scriptsize SCIPY \normalsize \citep{Virtanen_et_al_2020} method \texttt{scipy.integrate.cumulative\_trapezoid()}. This allows us to estimate the deviation of the simulated J-factor value calculated over the given interval from the fitted smoothed profile, thus estimating the contribution of any small scale profile irregularities in the J-factor profiles to the simulated J-factor value. This is important to calculate given the simulation could cause spurious substructure in the 1D profiles. However, the significant tidal disruption of Sgr also could cause substructure not accurately modelled by conventional smooth profiles. Over a range $-20^\circ<\beta_\odot<20^\circ$, the percentage difference between the total integrated J-factor value of the scaled $\beta_\odot$ J-factor profile and the integrated value of the fitted Moffat distribution (both depicted in Figure~\ref{fig:DM_J_factor_Beta_profile_normalized}) was calculated as $1.34\%$. Similarly, the percentage difference in integrated J-factor values between the scaled $\Lambda_\odot$ J-factor profile and fitted Moffat distribution (depicted in Figure~\ref{fig:DM_J_factor_Lambda_profile_normalized}) was calculated as $3.0\%$ over the range $-20^\circ<\Lambda_\odot<20^\circ$. These values imply a similarly low contribution from small scale J-factor distribution variance within these intervals to the J-factor values reported in this study, in spite of the marked small scale variation exhibited by these profiles.
\\[10pt]
As is evident in Figures \ref{fig:DM_J_factor_Beta_profile} and \ref{fig:DM_J_factor_Lambda_profile}, whilst a point source approximation for the J-factor distribution of Sgr is clearly invalid, the J-factor distribution is highly concentrated with a significant fraction of the total emission situated within the core radius $\alpha_{\text{Sgr}} \simeq 3.7^\circ$ of Sgr reported in \citet{Majewski_et_al_2003}. Figures \ref{fig:Cumulative_J_factor_plot_Beta} and \ref{fig:Cumulative_J_factor_plot_Lambda} show the cumulative summed J-factor value as a function of $|\beta_\odot|$ and $|\Lambda_\odot|$, respectively, calculated by summing the individual particle J-factor values (in accordance with equation~\ref{EQ9}) within the given angular distance $r$ from the centre of the simulated Sgr halo. Note that when summing along each axis (for example $\beta_\odot$), the summation range along the alternate axis (for example $\Lambda_\odot$) is unrestricted. Also shown are the angular distances along each axis containing $50\%$ and $90\%$ of the total summed J-factor value within $20^\circ$ of the centre of the simulated Sgr halo. Along the $\beta_\odot$ axis, $50\%$ ($90\%$) of the total cumulative J-factor emission within $|\beta_\odot|=20^\circ$ occurs within an angular distance of $|\beta_\odot|=1.7^\circ$ ($|\beta_\odot|=4.8^\circ$), whilst some $\simeq 83\%$ of the total cumulative J-factor emission is located within the Sgr core radius. Along the $\Lambda_\odot$ axis, the core radius of Sgr contains $62\%$ of the total cumulative J-factor emission within $|\Lambda_\odot|=20^\circ$; $90\%$ of the total J-factor emission within $|\Lambda_\odot|=20^\circ$ is located within an angular distance of $|\Lambda_\odot|=8.6^\circ$.
\begin{figure}
    \centering
    \includegraphics[width=\columnwidth]{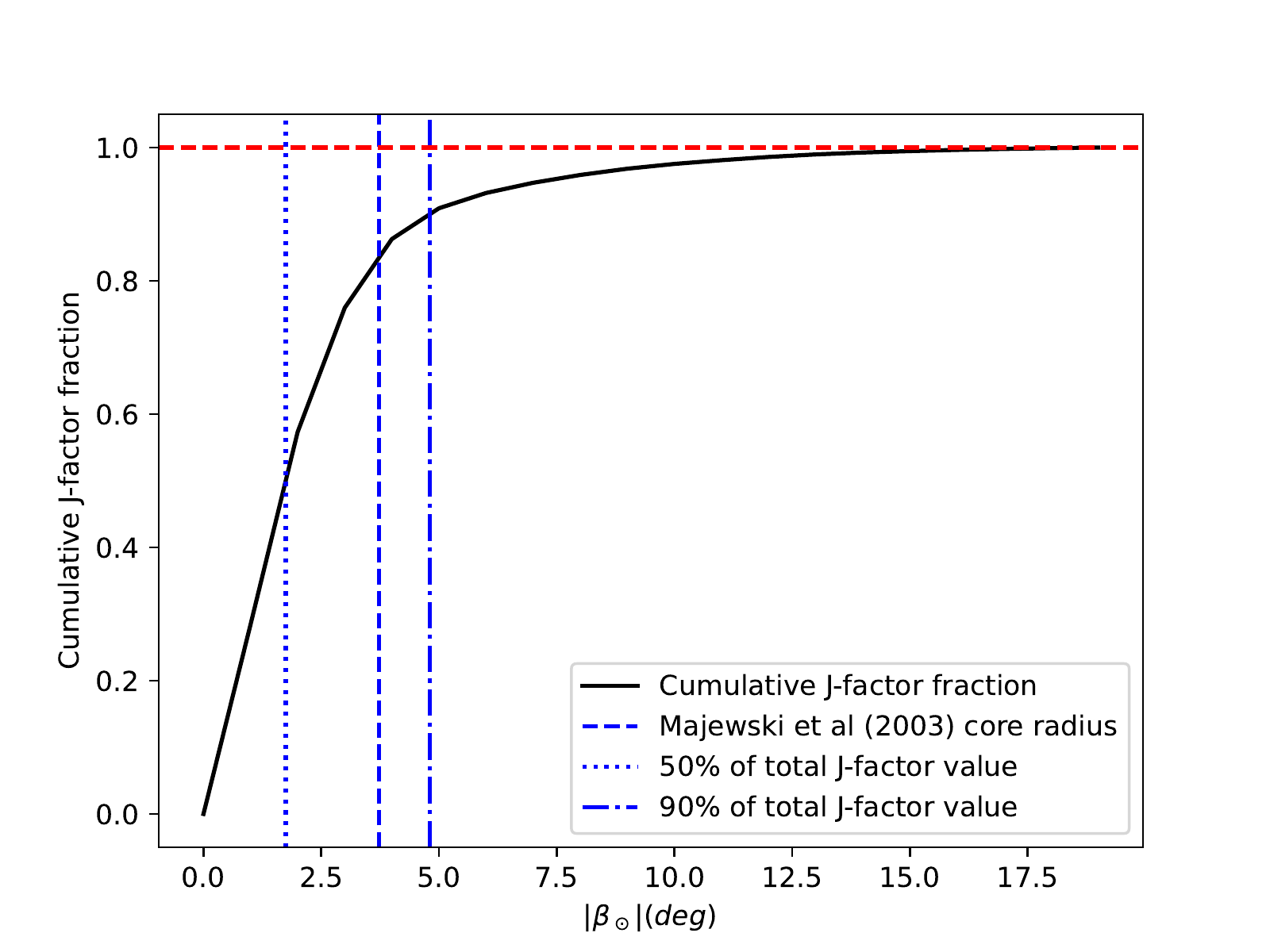}
    \caption{The cumulative J-factor fraction (within $|\beta_\odot|=20^\circ$) as a function of $|\beta_\odot|$. As discussed in section~\ref{sec:1D_J_factor_profiles_results}, the core radius of Sgr reported in \citet{Majewski_et_al_2003} contains $83\%$ of the cumulative J-factor emission.}
    \label{fig:Cumulative_J_factor_plot_Beta}
\end{figure}
\begin{figure}
    \centering
    \includegraphics[width=\columnwidth]{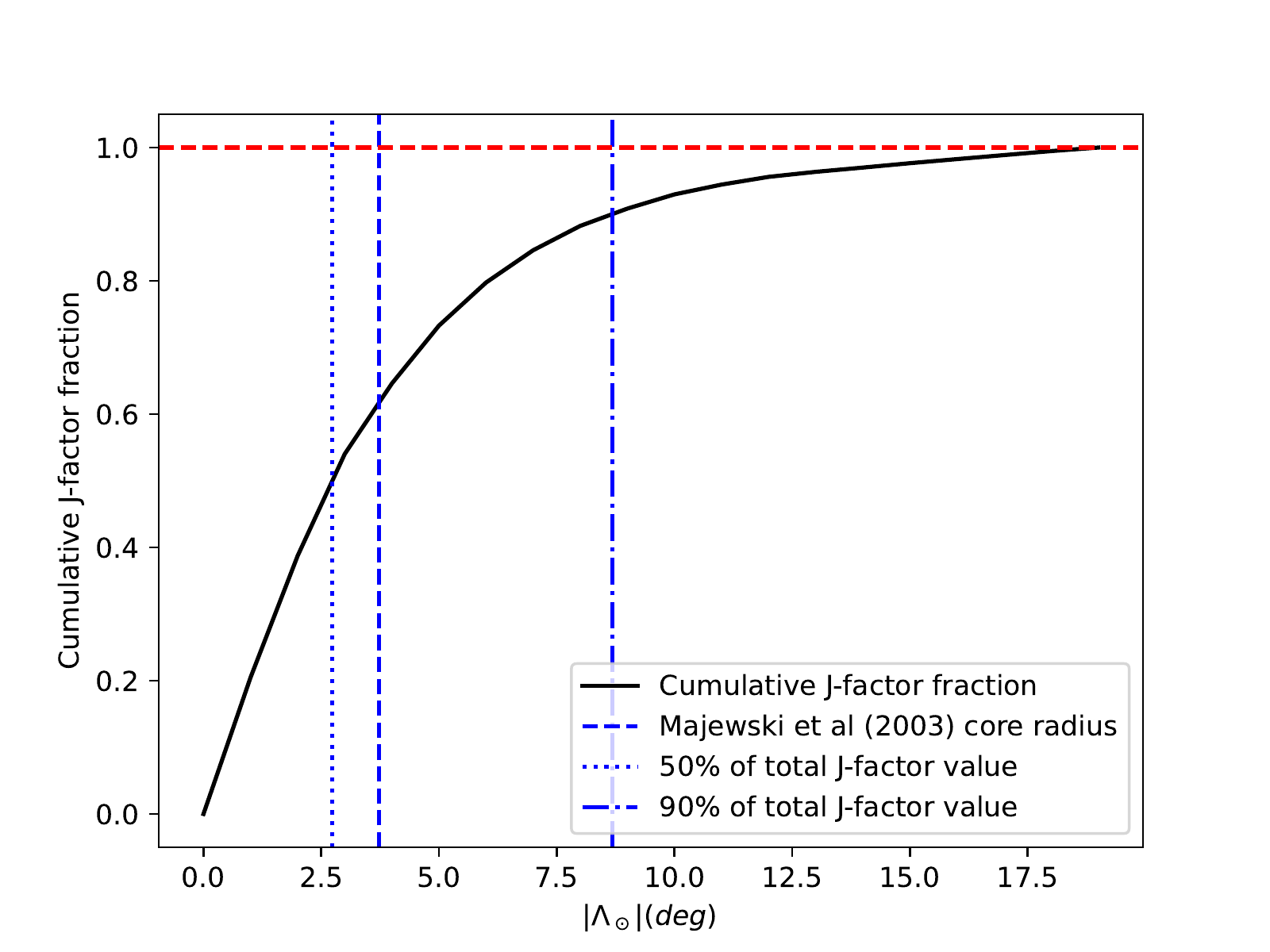}
    \caption{The cumulative J-factor fraction (within $|\Lambda_\odot|=20^\circ$) as a function of $|\Lambda_\odot|$. As discussed in section~\ref{sec:1D_J_factor_profiles_results}, the core radius of Sgr contains $62\%$ of the cumulative emission over this interval.}
    \label{fig:Cumulative_J_factor_plot_Lambda}
\end{figure}
\subsection{Projected $\rho_s$ and $\rho_{DM}^2$ profiles}
\label{sec:Rho_profiles_section}
\begin{figure}
    \centering
    \includegraphics[width=\columnwidth]{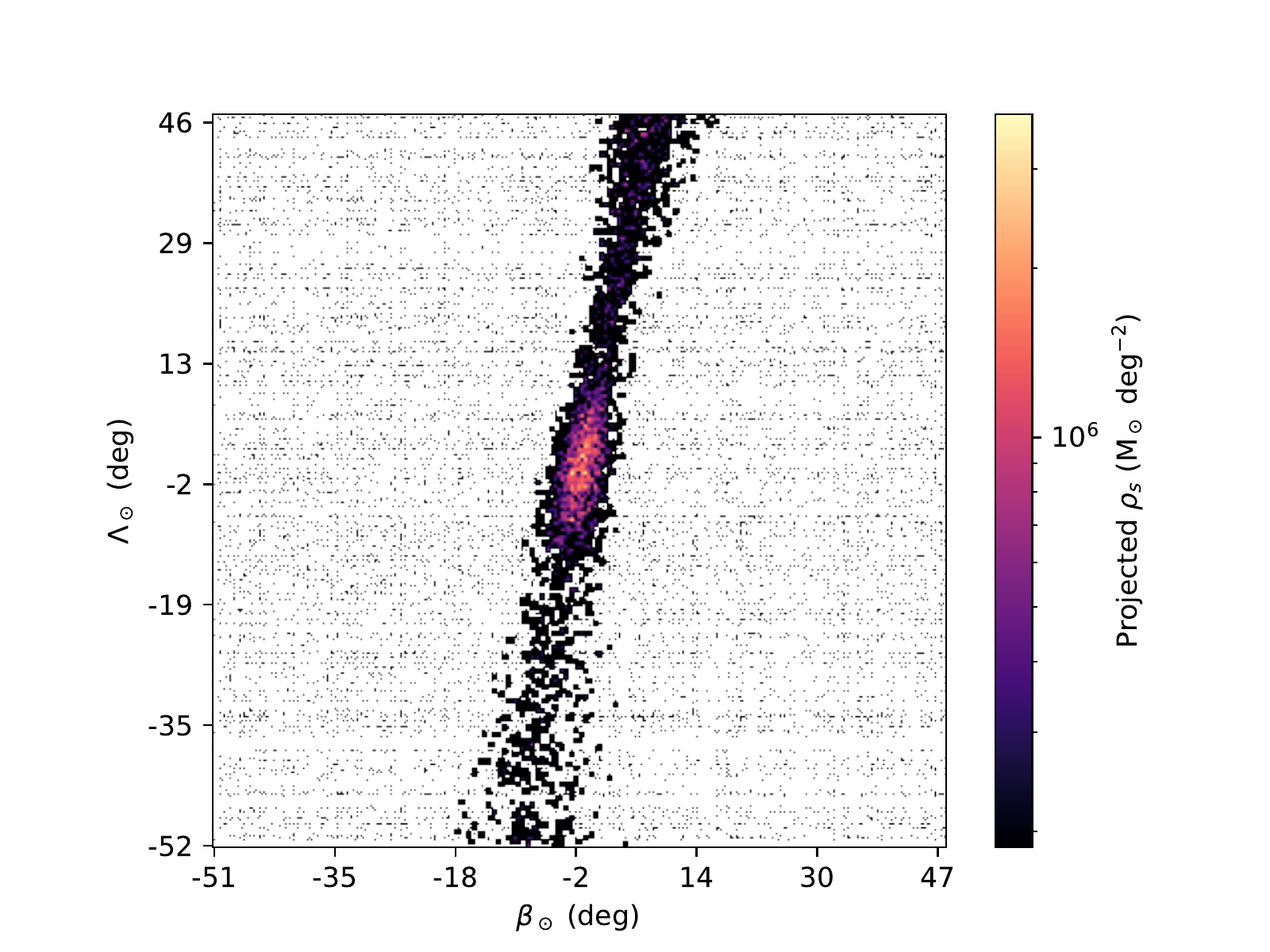}
    \caption{The projected $\rho_s$ distribution for Sgr, depicted in the coordinate system of \citet{Belokurov_et_al_2014}, in the angular units of M$_\odot$ deg$^{-2}$. The adopted pixel size in this figure is $\alpha^2=0.21$ square degrees. The projected $\rho_{s}$ peak is located at $(\beta_\odot,\Lambda_{\odot}) \simeq (-2.4,-0.9)^\circ$, an offset of approximately $1.4^\circ$ from the projected dark matter $\rho_{DM}^2$ peak.} 
    \label{fig:Rho_sim_stars_ROI_angular_units}
\end{figure}

\begin{figure}
    \centering
    \includegraphics[width=\columnwidth]{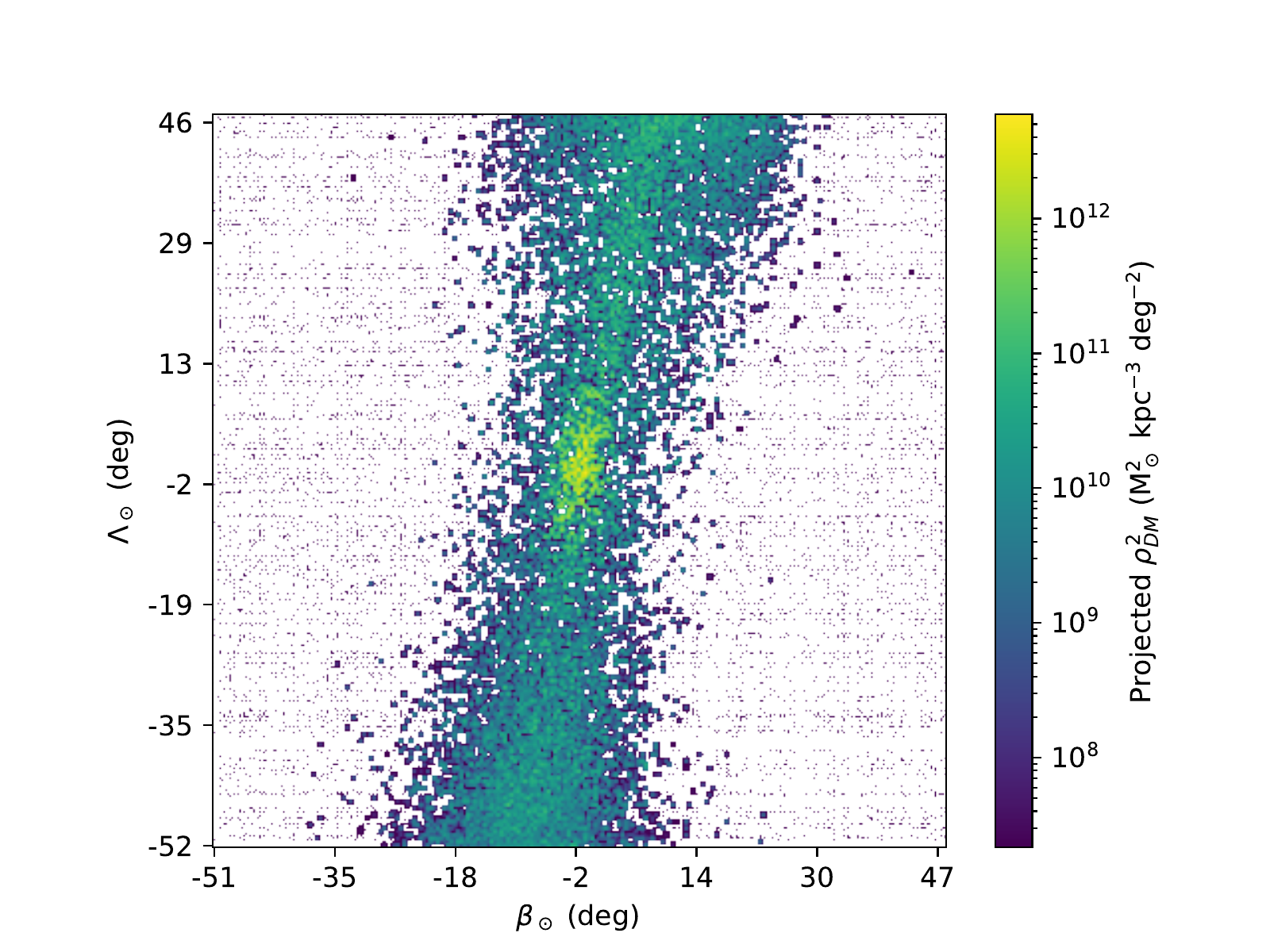}
    \caption{The projected $\rho_{DM}^2$ distribution for Sgr in the coordinate system of \citet{Belokurov_et_al_2014}, in the angular units of M$_\odot^2$ kpc$^{-3}$ deg$^{-2}$. The adopted pixel size in this figure is $\alpha^2=0.21$ square degrees. The increased dark matter density within the core radius of Sgr is clearly visible, with the projected $\rho_{DM}^2$ peak located at $(\beta_\odot,\Lambda_{\odot}) \simeq (-1.0,-1.3)^\circ$.}
    \label{fig:Rho_squared_DM_ROI_angular_units}
\end{figure}

Given that explaining the observed $\gamma$-ray emission from Sgr with the derived J-factor distribution is inconsistent with existing constraints on DM annihilation cross sections, this section describes the distributions of simulated stellar density ($\rho_s$) and the square of DM density ($\rho_{DM}^2$) to further inform observational analyses of Sgr. The expected $\gamma$-ray emission flux from stellar-associated sources is proportional to the projected stellar density \citet{Crocker_and_Macias_et_al_2022}, whilst the flux of $\gamma$-ray emission from DM annihilation is proportional to the projected J-factor distribution (see section~\ref{sec:J_factor_calculation_method}). This section facilitates a comparison of the morphology of the expected emission from these two sources. However, it should be noted that the actual $\gamma$-ray emission flux from these sources is highly dependent on the assumed $\gamma$-ray emission model; accordingly, the relative magnitudes of the J-factor and $\rho_s$ distributions do not provide an accurate indication of the primary dominant $\gamma$-ray emission mechanism expected for Sgr.
\\[10pt]
Figure~\ref{fig:Rho_sim_stars_ROI_angular_units} displays the projected $\rho_s$ distribution at the location of the Sagittarius core, calculated for each pixel in accordance with equation~\ref{eq:Simulated_Stellar_projected_mass_density} by dividing the sum of particle mass in the pixel by the pixel size ($\alpha^2=0.21$ square degrees). As per the 2D projected J-factor distributions discussed in section~\ref{sec:J_factor_calculation_method}, all figures in this section are depicted in the ($\beta_\odot,\Lambda_\odot$) coordinate system of \citet{Belokurov_et_al_2014} and were produced utilising the HEALPix pixelization scheme \citet{Gorski_et_al_2005} to sum the relevant particle quantities in spatial pixels.
\\[10pt]
As discussed in section~\ref{sec:J_factor_calculation_method}, the Sgr $\rho_{DM}^2$ (and DM density) distribution is difficult to model with stellar tracers due to the ongoing tidal disruption of the Sgr halo. The $\rho_{DM}^2$ distribution also is equal to the absolute magnitude of the Sgr J-factor definition, as detailed in section~\ref{sec:J_factor_calculation_method}, which allows comparison to other studies of indirect dark matter annihilation utilising absolute magnitude J-factor definitions (e.g \citealt{Stoehr_et_al_2003}). Accordingly, in this appendix we also present the projected Sgr $\rho_{DM}^2$ distribution to inform future studies attempting to calculate the DM density distribution or the J-factor distribution of Sgr. Figure~\ref{fig:Rho_squared_DM_ROI_angular_units} shows this projected $\rho_{DM}^2$ distribution for Sgr. The displayed projected $\rho_{DM}^2$ value for each pixel was calculated by dividing the result of equation~\ref{EQ8} for each pixel by the (constant) pixel size ($\alpha^2=0.21$ square degrees).
\\[10pt]
A quantity of potential interest to observational analyses of Sgr is the projected offset in maximum density between the simulated stellar and dark matter populations. To calculate this offset, the projected $\rho_{s}^2$ (not $\rho_{s}$) and $\rho_{DM}^2$ distributions were smoothed, such that the value of a pixel $b$ in the smoothed distribution was the mean value of the density distribution in pixel $b$ and the immediate 8 surrounding pixels. This smoothing minimized the impact of small-scale density fluctuations on determining the position of maximum density, as did the use of the $\rho_{s}^2$ instead of the $\rho_{s}$ distribution. The peak of the $\rho_{s}$ distribution is necessarily coincident with the $\rho_{s}^2$ distribution peak.
\\[10pt]
The pixel corresponding to the projected maximum squared density was then located for both simulated stellar and dark matter populations. In the coordinate system of \citet{Belokurov_et_al_2014}, these pixels were located at $(\beta_\odot,\Lambda_{\odot}) = (-1.02,-1.31)^\circ$ and $(\beta_\odot,\Lambda_{\odot}) = (-2.40,-0.85)^\circ$ for the dark matter and simulated stellar particle populations, respectively (see section~\ref{sec:J_factor_calculation_method} for a detailed explanation of this coordinate system). The projected offset between these pixel positions was thus $1.44^\circ$.
\\[10pt]
In their $\gamma$-ray analysis employing a Sgr stellar template, \citet{Crocker_and_Macias_et_al_2022} found moderate evidence ($4.5\sigma$ statistical significance) for a shift in the best-fit position of the template 180$^\circ$ away from the dwarf galaxy's travel direction ($\sim 4^\circ$ towards the Galactic south). This is evidence for an offset between the peak of the $\gamma$-ray emission and the centre of Sgr defined by its stars. The much smaller offset, found here, between the simulated DM population and the simulated stellar population of Sgr does not seem to provide an explanation of the offset tentatively found by \citet{Crocker_and_Macias_et_al_2022}. Furthermore, the offset between the simulated stellar and dark matter density peaks we find is in the wrong direction to explain the offset found by \citet{Crocker_and_Macias_et_al_2022}. This in turn tentatively indicates that the observed $\gamma$-ray emission is not due to $\gamma$-ray emission from DM annihilation in Sgr, consistent with \citet{Crocker_and_Macias_et_al_2022}.
\\[10pt]
As aforementioned, comparative analysis of the J-factor and $\rho_s$ distributions provides valuable information on the comparative morphology of the expected $\gamma$-ray emission from DM annihilation and stellar-associated sources, whilst the $\rho_{DM}^2$ distribution provides valuable insight on the shape of the Sgr DM halo. Accordingly, following the methodology outlined in section~\ref{sec:1D_J_factor_profiles_results}, we derived projected 1D $\rho_s$ and $\rho^2_{DM}$ profiles from the 2D distributions respectively depicted in Figure \ref{fig:Rho_sim_stars_ROI_angular_units} and Figure~\ref{fig:Rho_squared_DM_ROI_angular_units}, as functions of $\Lambda_\odot$ and $\beta_\odot$. These distributions were then scaled (such that their maximum value was $1.0$) and fitted with distribution functions to measure their width and facilitate comparison to the 1D projected J-factor profiles discussed in section~\ref{sec:1D_J_factor_profiles_results}. As per the 1D projected J-factor profiles discussed in section~\ref{sec:1D_J_factor_profiles_results}, the choice of fitted function was selected by minimizing the non-linear least squares optimization detailed in equation~\ref{eq:Opt_function}. The functions were again fitted to the 1D profiles using a Levenberng-Marquardat \citep{Levenberg_1944,Marquardt_1963} non-linear least squares algorithm implemented in \scriptsize ASTROPY \normalsize (\citealt{astropy:2013,astropy:2018,astropy:2022}, in the case of the Moffat, Gaussian or Voigt functions) or \scriptsize SCIPY \normalsize (\citealt{Virtanen_et_al_2020}, in the case of the offset Gaussian function). The offset Gaussian function, which best fit the 1D projected $\rho_s$ and $\rho^2_{DM}$ distributions as functions of $\Lambda_\odot$, is a Gaussian with an additional nonzero constant of the form
\begin{equation}
    f(x) = \frac{a}{\sigma \sqrt{2\pi}}exp{\left(\frac{-(x-\mu)^2}{2\sigma^2}\right)} + c
    \label{EQ:Offset_Gaussian}
\end{equation}
\\[10pt]
The parameters of the distribution functions fitted to these scaled 1D projected $\rho_s$ and $\rho^2_{DM}$ profiles are detailed in Table~\ref{tab:Fits_summary_table}, along with their FWHM values. Note the considerably larger FWHM values measured for the $\Lambda_\odot$ profiles, with poorer fits, resulting from the extensive tidal disruption of Sgr. The FWHM measurements for the projected 1D $\rho_s$ profiles profiles demonstrate the extended nature of any expected $\gamma$-ray emission from stellar-associated sources in Sgr, with the measured FWHM values for these profiles broadly consistent with the extended emission distribution detected with $8.1\ \sigma$ significance in \citet{Crocker_and_Macias_et_al_2022}. Together, the measured FWHM values for the six profiles listed in Table~\ref{tab:Fits_summary_table} suggest against assuming a point-source emission profile during the analysis of any $\gamma$-ray emission from Sgr. Extending the $\rho^2_{DM}$, $\rho_s$ and J-factor distributions to analyse the Sagittarius Stream will be the subject of Paper II.
\section{Discussion: Comparison of J-factor magnitude with past studies of the Sagittarius Dwarf}
\label{sec:Discussion}
Only a few studies have estimated the J-factor value of the Sgr dark matter component; these estimates are consistently larger than the J-factor value $J_{\text{Sgr}}=1.48\times 10^{10}$  M$_\odot^2$ kpc$^{-5}$ calculated in this work. We will explore several of these works and compare them to our own calculations in turn.
\\[10pt]
Firstly, \citet{Abramowski_et_al_2014} calculate a J-factor value of $J_{\text{Sgr},\text{Abramowski}} \simeq 2.9\times 10^{12}$ M$_\odot^2$ kpc$^{-5}$, estimating the dark matter density profile utilising a NFW profile fit to stellar velocity measurements. They then utilise a maximum-likelihood process to determine the most probable contribution of this dark matter distribution toward $\gamma$-ray counts observed with the H.E.S.S. telescope. \citet{Abramowski_et_al_2014} marginally detect Sgr with a $2.05\ \sigma$ significance, however conclude that $\gamma$-ray counts due to the dark matter population of Sgr (as opposed to astrophysical sources) are likely negligible.
\\[10pt]
Secondly, \citet{Viana_et_al_2012} fit a NFW profile to Sloan Digital Sky Survey (SDSS) stellar velocity dispersion data. This profile is integrated within an integration area of $\Delta \Omega = 2\times10^{-5}\ \text{sr} = 0.07\ \text{deg}^2$ to derive a J-factor value of $J_{\text{Sgr},\text{Viana}} = 2.0\times10^{16}$  M$_\odot^2$ kpc$^{-5}$, with uncertainties of a factor of 2.
\begin{figure*}
    \centering
    \includegraphics[width=2\columnwidth]{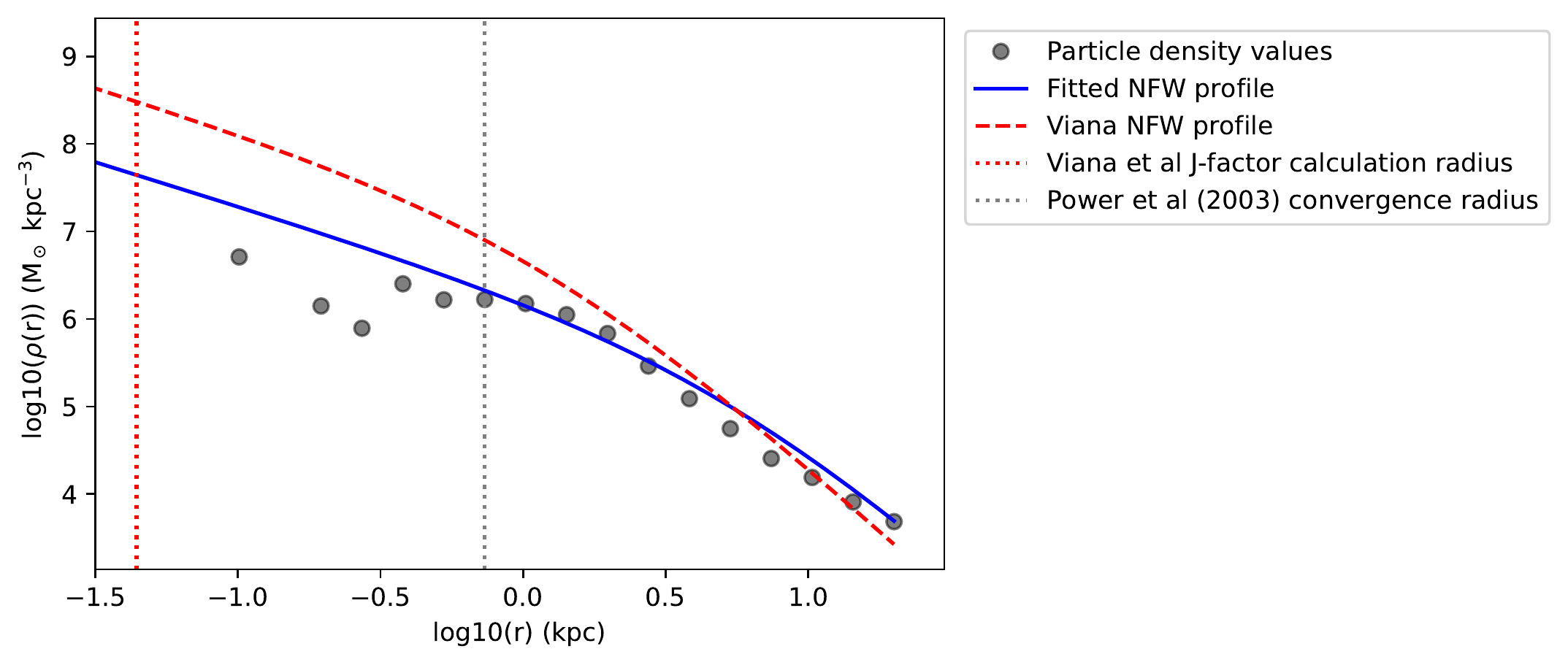}
    \caption{The simulated particle density profile as a function of the radius $r$ from the centre of the simulated Sgr DM halo, with the fitted NFW profile to the simulation particle densities (in blue) and the NFW profile fitted by \citet{Viana_et_al_2012} to stellar velocity dispersion data (in red). The \citet{Power_et_al_2003} convergence radius $r_c=0.73$ kpc of the simulated particle halo is indicated; the radial density profile internal to this radius was not utilised when fitting the NFW profile indicated in blue. Also indicated is the radius $r=0.044$ kpc, corresponding to a projected circular area of $2\times10^{-5}$ sr, utilised to calculate the J-factor in \citet{Viana_et_al_2012}. Clearly, the fitted NFW profile will provide a significantly lower density estimate at this radius than the NFW profile detailed in \citet{Viana_et_al_2012}, but still overestimates the radial density profile at small radii.}
    \label{fig:Viana_comparison_plot}
\end{figure*}
\\[10pt]
 Given the lack of detection of higher predicted dark matter J-factor fluxes in \citet{Viana_et_al_2012,Abramowski_et_al_2014} and significantly stronger observed $\gamma$-ray emission from MSP candidates in Sgr \citep{Crocker_and_Macias_et_al_2022}, the derived J-factor value for Sgr in this study is unlikely to improve current constraints on $\gamma$-ray emission from the Sgr dark matter distribution.
\\[10pt]
To investigate the discrepancy between this study and the larger J-factor value for Sgr reported in \citet{Viana_et_al_2012} we fitted a NFW profile of the form \citep{Navarro_et_al_1997}
\begin{equation}
    \rho(r) = \frac{\delta_c}{(r/r_s)(1+r/r_s)^2}
    \label{NFW_density_profile}
\end{equation}
to the simulated radial DM mass density profile (calculated as the mass density of concentric shells), where $\rho(r)\ \text{M}_{\odot}\text{kpc}^{-3}$ is the density at a radial distance of $r$ kpc, $\delta_c$ is the characteristic density of the halo and $r_s$ is the scale radius of the halo. As a precaution, only particles at a radius of greater than the \citet{Power_et_al_2003} convergence radius $r_c=0.73$ kpc from the centre of Sgr were utilised in the fitting procedure to avoid potentially unrealistic density profiles due to simulation artefacts. The fitted NFW density profile and simulated particle density profile are detailed in Figure~\ref{fig:Viana_comparison_plot} as a function of the radius $r$ from the centre of the simulated DM halo. The radial density profile at smaller radii is significantly overestimated by the fitted NFW profile. Thus, relaxing this fitting constraint and fitting the radial density profile at smaller radii would result in shallower inner NFW profile slope and inner profile density values, with corresponding downward revision of any J-factor value estimated from the fitted NFW profile. The fitted NFW profile parameters and the equivalent NFW profile parameters utilised in \citet{Viana_et_al_2012} are detailed in table~\ref{tab:NFW_parameters_table}.
\begin{table*}
    \centering
    \caption{The parameters of the NFW profiles either utilised in \citet{Viana_et_al_2012} or fitted to the simulated particle density profile in this study, with the J-factor values calculated from these profiles utilising the analytic J-factor definition detailed in equation~\ref{Sigma_M_squared_dist} (see section~\ref{sec:Discussion}). The projected area of integration was $\Delta \Omega=2\times10^{-5}$ sr in both cases. The errors quoted are one standard deviation on the fitted parameters.}
    \begin{tabular}{|c|c|c|c|}
        \hline
        Parameters & $r_s$ (kpc) & $\delta_c$ ($\text{M}_{\odot}\text{kpc}^{-3}$) & J-factor value (M$_\odot^2$ kpc$^{-5}$)  \\
        \hline
        \citet{Viana_et_al_2012} & $1.3$ & $1.1\times10^7$ & $1.1\times10^{16}$\\
        Fitted & $6\pm2$ & $(3\pm2)\times10^4$ &  $2.5\times10^{14}$\\
        \hline
    \end{tabular}
    \label{tab:NFW_parameters_table}
\end{table*}
\\[10pt]
A J-factor value was calculated from the fitted NFW profile utilising the equation
\begin{equation}
    J(R) = 2\int_R^{\sqrt{r_d^2+R^2}} \frac{r\rho^2(r)}{\sqrt{r^2-R^2}}\ \text{d}r
    \label{Sigma_M_squared_dist}
\end{equation}
where $\rho(r)$ was the NFW density profile given by equation~\ref{NFW_density_profile} with `Fitted' parameters detailed in table~\ref{tab:NFW_parameters_table}. $r_d$ is a dark matter density profile truncation radius; we adopt the radius of $r_d=4\ \text{kpc}$ assumed by \citet{Viana_et_al_2012} for the purposes of comparison, though variation of this value causes negligible change in our results. Following \citet{Viana_et_al_2012}, a radius of $R=0.044$ kpc was utilised, equivalent to a projected circular area of $\Delta \Omega = 2\times10^{-5}\ \text{sr}$ at the distance of Sgr. This corresponds to a projected angular radius of $0.1^\circ$. The resulting J-factor value calculated with equation~\ref{Sigma_M_squared_dist} was $J_{\text{Sgr},\text{NFW}} = 2.5\times10^{14}$  M$_\odot^2$ kpc$^{-5}$. It is important to note that this J-factor value is independent of the assumed particle-based J-factor definition.
\\[10pt]
Repeating the calculation detailed in section~\ref{sec:Predicted_J_factor_magnitude_result}, the lower limit on the WIMP particle annihilation cross section required to explain the $\gamma$-ray photon flux attributed to Sgr in \citet{Crocker_and_Macias_et_al_2022}, assuming this NFW J-factor value, is also inconsistent with current constraints \citep{Abazajian_et_al_2020,Evans_et_al_2023} for WIMP masses $\gtrsim 10$ GeV.
\\[10pt]
Calculating a J-factor value using equation~\ref{Sigma_M_squared_dist} and an identical radius from the NFW profile $\rho(r)$ defined by the parameters of \citet{Viana_et_al_2012} yielded a J-factor value of $1.1\times10^{16}$ M$_\odot^2$ kpc$^{-5}$. This is consistent within errors of the value $J_{\text{Sgr},\text{Viana}} = 2\times10^{16}$ M$_\odot^2$ kpc$^{-5}$ reported in \citet{Viana_et_al_2012}, though our calculation uses a different J-factor definition. This in turn implies that the difference between the total J-factor value calculated from the fitted NFW profile $J_{\text{Sgr},\text{NFW}} = 2.5\times10^{14}$ M$_\odot^2$ kpc$^{-5}$ and the result of \citet{Viana_et_al_2012} can be partially explained by the relatively lower density of our fitted NFW density profile at small radii, as evident in Figure~\ref{fig:Viana_comparison_plot} (noting that the J-factor scales as density squared).
\\[10pt]
To further explore the effect of differing J-factor definitions on the calculated J-factor value for Sgr, the value of the J-factor as a function of radius $r$ from the centre of the simulated DM halo was calculated utilising both the analytic definition detailed in equation~\ref{Sigma_M_squared_dist} and the J-factor calculated from the pixel summation process detailed in equation~\ref{EQ9}. The analytic J-factor values were calculated from a radius of $r\geq 0.001$ kpc, comparable to the physical pixel width at the distance of Sgr, namely $0.0035$ kpc. These are detailed in Figure~\ref{fig:J_factor_values_comparison}.
\begin{figure*}
    \centering
    \includegraphics[width=2\columnwidth]{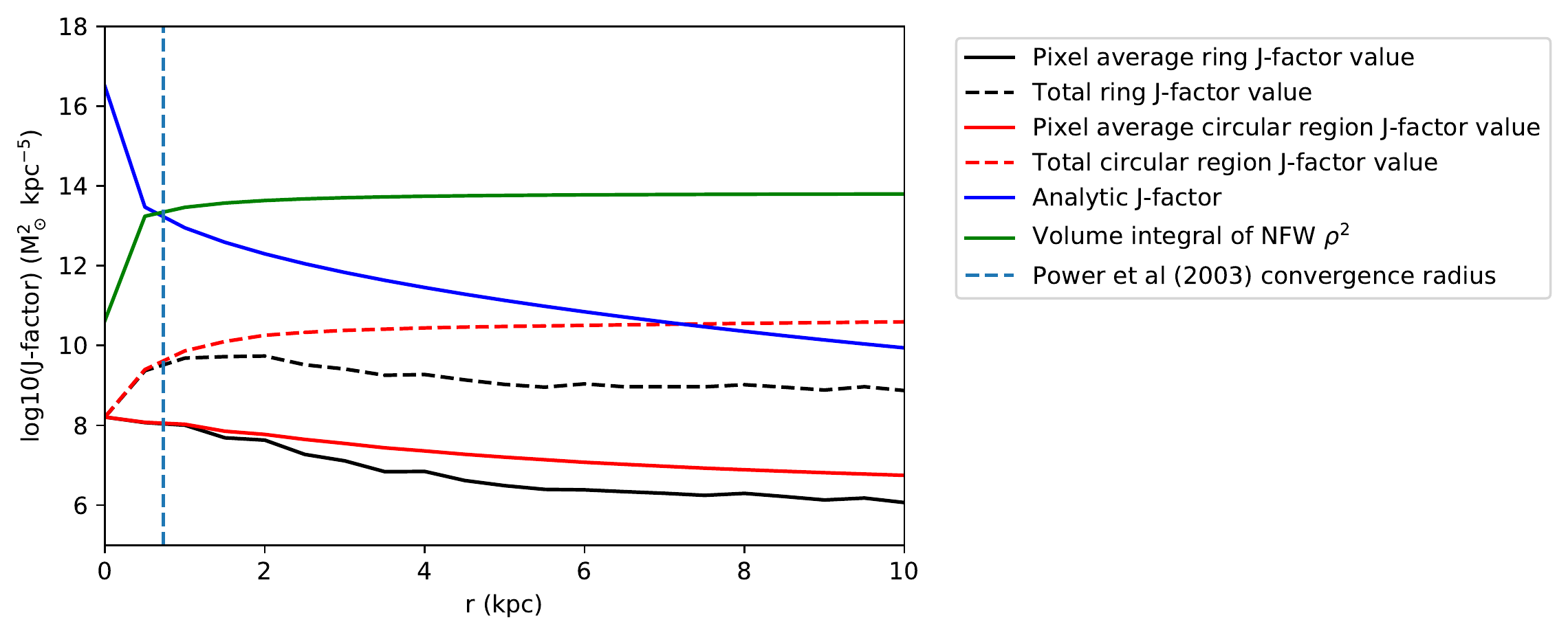}
    \caption{J-factor values for Sgr, as a function of radius $r$ from the centre of the simulated DM halo, computed utilising various definitions. `Ring' J-factor values are computed utilising equation~\ref{EQ9}, summing particles along the line of sight in a thin ring of radius $r$ kpc centred on Sgr. `Circular region' J-factor values are also computed utilising equation~\ref{EQ9}, summing over all particles along the line of sight within a circular region of radius $r$ kpc centred on Sgr. The analytic J-factor value is computed utilising equation~\ref{Sigma_M_squared_dist} from the fitted NFW profile depicted in Figure~\ref{fig:J_factor_values_comparison}. Also shown in green  is the volume integral of the fitted NFW $\rho_{DM}^2$ profile within a spherical volume of radius $r$ kpc centred on Sgr, computed utilising equation~\ref{Volume_integral}.}
    \label{fig:J_factor_values_comparison}
\end{figure*}
The `circular region' J-factor values (indicated in solid red) are calculated utilising equation~\ref{EQ9}, summing pixel contributions within a circular-based conical area of radius $r$ at the distance of Sgr. The `ring' J-factor values (indicated in solid black) are again calculated utilising equation~\ref{EQ9} but summing pixels along the circumference of a thin ring of radius $r$. As would be expected, this approaches the circular region J-factor value for small radii. The average J-factor value per pixel for both the circular summation regions (in red dashes) and the ring summation regions (in black dashes) are also displayed. In blue on Figure~\ref{fig:J_factor_values_comparison} is the J-factor value calculated with equation~\ref{Sigma_M_squared_dist} from the NFW density profile $\rho(r)$ fitted to the simulated dark matter particle density profile (Figure~\ref{fig:Viana_comparison_plot}), implemented utilising a modified variant of the \texttt{\_surfaceDensity} method from the \scriptsize COLOSSUS \normalsize package \citep{Diemer_2018}. Lastly, in green the volume integral of density squared
\begin{equation}
     \int \rho^2\ \text{d}V = 4\pi\int_0^r \rho^2(r)r^2\ \text{d}r
     \label{Volume_integral}
\end{equation}
is displayed, again calculated utilising the NFW density profile $\rho(r)$ fitted to the dark matter particle density profile. Note that, in contrast to this study, the volume integral of density squared is also defined as the J-factor value in \citet{Stoehr_et_al_2003}.
\\[10pt]
As previously detailed, the `total circular region J-factor value' detailed in Figure \ref{fig:J_factor_values_comparison}, as calculated with equation~\ref{EQ9}, was selected to calculate the Sgr J-factor value in this study. The following considerations motivated this selection over other definitions. Firstly, utilising a simulation based particle summation J-factor definition likely more accurately accounts for the extended, irregular shape of the Sgr dark matter halo due to tidal disruption \citep{Lokas_et_al_2010}. Calculating the J-factor value directly from the simulated particle distribution rather than fitting stellar tracers or a profile also avoids assumptions of dynamical equilibrium or spherical symmetry, which are clearly invalid in the case of Sgr. Given the extrapolated NFW profile clearly overestimates the particle density profile in the inner region of the simulated Sgr halo, the particle-based J-factor value is again a conservative lower limit accounting for potential overestimation due to an overestimated fitted NFW profile density. However, some of the density discrepancy between the fitted NFW and particle based density profiles at small radii \textit{may} result from \textit{potentially} insufficient simulation resolution at small radii. In contrast, the J-factor value calculated from the fitted NFW profile likely exceeds the true J-factor value for Sgr, given the overestimate of the fitted NFW profile to the simulated particle density profile at small radii, in addition the divergent behaviour of the NFW density profile and analytic J-factor definition at small radii.
\\[10pt]
Secondly, in accordance with equation~\ref{EQ9} it is clear that the J-factor definition should exhibit a similar behaviour as a function of radius as the volume integral of density squared, particularly for small radii, as all particles are at a similar heliocentric distance. The pixel averaged ring and pixel averaged circular definitions do not follow this behaviour, nor does the analytic definition detailed in equation~\ref{Sigma_M_squared_dist} which shows a nonphysical divergence at small radii. Lastly, the ring J-factor value does not serve to calculate the total integrated J-factor value within an extended structure, as required by this study.
\\[10pt]
Consideration of these different J-factor definitions also provides additional insight into potential causes for the discrepancy between the Sagittarius Dwarf J-factor value calculated in this study and prior results, such as that reported in \citet{Viana_et_al_2012}. The analytic J-factor definition detailed in equation~\ref{Sigma_M_squared_dist}, which reproduces the result of \citet{Viana_et_al_2012} (within the margin of error) at a radius of $r=0.044$ kpc from the centre of the simulated DM halo, is clearly divergent at such small radii when calculated utilising a fitted NFW profile. This is a source of discrepancy between the J-factor value calculated from simulated particle data using equation~\ref{EQ9} and the J-factor value for Sgr calculated in \citet{Viana_et_al_2012} (and other prior works) in addition to the aforementioned discrepancy resulting from the differing density profiles.
\\[10pt]
Similarly to \citet{Viana_et_al_2012}, an estimate of the dark matter density derived through Jeans analysis is utilised by \citet{Evans_et_al_2023} to compute a J-factor value of $J_{\text{Sgr},\text{Evans}}=9.13\times10^{12}$ M$_\odot^2$ kpc$^{-5}$ ($10^{19.6}\ \text{GeV}\ \text{cm}^{-5}$). The analysis of \citet{Evans_et_al_2023} does not explore the possibility of extended emission beyond a radius of $2^\circ$ from the centre of the Sgr and assumes a NFW profile of small scale radius ($r_s=1$ kpc). We have shown that the DM density distribution of Sgr is significantly extended and find the NFW profile fitted to the Sgr DM density profile has a significantly larger scale radius than assumed in \citet{Evans_et_al_2023} and \citet{Viana_et_al_2012}. The simulated DM radial density profile (displayed in Figure~\ref{fig:Viana_comparison_plot}) indicates a significantly lower core DM density than utilised in \citet{Evans_et_al_2023} and \citet{Viana_et_al_2012} is appropriate for Sgr. This, in addition to the significant tidal disruption of the simulated Sgr halo over a large radial range, reinforces the conclusion of \citet{Evans_et_al_2023} that their DM density profile derived using Jeans analysis likely significantly overestimates the J-factor value of Sgr.
\\[10pt]
Computing the J-factor value of Sgr utilising the simulated particle density profile, as detailed in this study, results in a significantly lower J-factor value for Sgr than found in \citet{Evans_et_al_2023} and implies a DM cross-section incompatible with the DM mass/annihilation cross section constraints illustrated in \citet[Figure 7]{Evans_et_al_2023}, further demonstrating that the Sgr halo should not be utilised for indirect DM detection searches.
\section{Summary and Conclusions}
\label{sec:Summary_and_conclusions}
A N-body/hydrodynamic simulation of the infall and tidal disruption of the Sagittarius Dwarf Galaxy (Sgr) around the Milky Way was utilised to investigate the expected integrated J-factor value of Sgr and explore the morphological characteristics of the projected J-factor distribution. The simulation of \citet{Tepper_Garcia_and_Bland_Hawthorn_2018} was utilised to produce J-factor distributions for the dark matter population of Sgr through a summation of particle density, distance and mass values in line of sight pixels. Utilising this methodology provides a more accurate model of the Sgr DM halo than derivations utilising stellar tracers through more accurately accounting for the strong tidal disruption of the Sgr DM halo. The markedly extended nature of the J-factor distributions imply the extended nature of any $\gamma$-ray source associated with the Sgr DM halo.
\\[10pt]
The J-factor value for Sgr, $J_{\text{Sgr}}=1.48\times 10^{10}$ M$_\odot^2$ kpc$^{-5}$ ($6.46\times10^{16}\ \text{GeV}\ \text{cm}^{-5}$), was calculated by summing the pixel J-factor values within the core radius of Sgr reported in \citet{Majewski_et_al_2003}. To explain the recently observed $\gamma$-ray emission from Sgr documented in \citet{Crocker_and_Macias_et_al_2022} with the derived J-factor value for Sgr would require a DM annihilation cross section incompatible with existing constraints. In conjunction with the recently observed $\gamma$-ray signal from MSP sources in \citet{Crocker_and_Macias_et_al_2022}, this J-factor value militates against the use of Sgr for indirect DM detection experiments. 
\\[10pt]
J-factor distributions were derived to provide insight into the morphology of potential dark matter annihilation signatures as a potential discriminant to astrophysical sources. The relative magnitude of the J-factor distributions at the centre of the Sgr halo and at the core radius of \citet{Majewski_et_al_2003} indicate that the DM J-factor distribution is insufficiently peaked to explain the morphology of the $\gamma$-ray emission from Sgr observed in \citet{Crocker_and_Macias_et_al_2022}, further indicating that DM annihilation is unable to explain the $\gamma$-ray emission observed in \citet{Crocker_and_Macias_et_al_2022}. Comparison of these distributions with fitted functions indicate low contamination from potentially spurious small-scale variation in the simulated J-factor distributions.
\\[10pt]
This calculated J-factor value for Sgr is lower than the J-factor values reported in prior studies (e.g \citealt{Viana_et_al_2012,Abramowski_et_al_2014}). To provide an indicative comparison between different J-factor definitions, analytic and various particle-based definitions were computed based on the simulated DM radial density profile. We show that the low J-factor value is due to differing density profiles for Sgr and a different J-factor definition, motivated by more accurate modelling of the tidally disrupted Sgr DM density profile.
\\[10pt]
Whilst the computed J-factor distribution militates against the use of Sgr in indirect DM detection searches, in future work we plan to determine the overall magnitude of the Sagittarius Stream J-factor and investigate its morphological characteristics with a view to probing Fermi-LAT data for potential $\gamma$-ray products of DM annihilation in the Sagittarius Stream.
\section*{Acknowledgements}
T.A.A.V and RMC would like to 
thank D. Mackey
for his contributions to the initial stages of this project.
T.A.A.V would like to 
acknowledge C. Blake, C. Power and A. Viana for helpful discussions. T.A.A.V. acknowledges the support of the Australian National University Research School of Astronomy and Astrophysics (RSAA) and the Centre for Astrophysics and Supercomputing at Swinburne University of Technology. 
RMC acknowledges support from the Australian Research Council through its \textit{Discovery Projects} funding scheme, awards DP190101258 and DP230101055.
O.M. was supported by the GRAPPA Prize Fellowship. T.T.G. acknowledges partial financial support from the Australian Research Council (ARC) through an Australian Laureate Fellowship awarded to J.~Bland-Hawthorn. This research was partially supported by the Australian Government through the Australian Research Council Centre of Excellence for Dark Matter Particle Physics (CDM, CE200100008). We acknowledge the facilities, and the scientific and technical assistance of the Sydney Informatics Hub (SIH) at the University of Sydney and, in particular, access to the high-performance computing facility Artemis and additional resources on the National Computational Infrastructure (NCI), which is supported by the Australian Government, through the University of Sydney’s Grand Challenge Program – the Astrophysics Grand Challenge: From Large to Small (CIs: G.~F.~Lewis and J.~Bland-Hawthorn). This research has made use of NASA’s Astrophysics Data System Bibliographic Services\footnote{https://ui.adsabs.harvard.edu}, the COLOSSUS\footnote{https://bdiemer.bitbucket.io/colossus/index.html} package \citep{Diemer_2018}, the Pyccl package, SciPy\footnote{https://scipy.org} \citep{Virtanen_et_al_2020} and Astropy\footnote{http://www.astropy.org}, a community-developed core Python package and an ecosystem of tools and resources for astronomy \citep{astropy:2013, astropy:2018, astropy:2022}.
\subsection*{Data Availability Statement}
No new data were generated or analysed in support of this research.



\bibliographystyle{mnras}
\bibliography{Sgr_paper_1} 

\begin{thebibliography}{}
\makeatletter
\relax
\def\mn@urlcharsother{\let\do\@makeother \do\$\do\&\do\#\do\^\do\_\do\%\do\~}
\def\mn@doi{\begingroup\mn@urlcharsother \@ifnextchar [ {\mn@doi@}
  {\mn@doi@[]}}
\def\mn@doi@[#1]#2{\def\@tempa{#1}\ifx\@tempa\@empty \href
  {http://dx.doi.org/#2} {doi:#2}\else \href {http://dx.doi.org/#2} {#1}\fi
  \endgroup}
\def\mn@eprint#1#2{\mn@eprint@#1:#2::\@nil}
\def\mn@eprint@arXiv#1{\href {http://arxiv.org/abs/#1} {{\tt arXiv:#1}}}
\def\mn@eprint@dblp#1{\href {http://dblp.uni-trier.de/rec/bibtex/#1.xml}
  {dblp:#1}}
\def\mn@eprint@#1:#2:#3:#4\@nil{\def\@tempa {#1}\def\@tempb {#2}\def\@tempc
  {#3}\ifx \@tempc \@empty \let \@tempc \@tempb \let \@tempb \@tempa \fi \ifx
  \@tempb \@empty \def\@tempb {arXiv}\fi \@ifundefined
  {mn@eprint@\@tempb}{\@tempb:\@tempc}{\expandafter \expandafter \csname
  mn@eprint@\@tempb\endcsname \expandafter{\@tempc}}}

\bibitem[\protect\citeauthoryear{{Abazajian}, {Horiuchi}, {Kaplinghat},
  {Keeley}  \& {Macias}}{{Abazajian} et~al.}{2020}]{Abazajian_et_al_2020}
{Abazajian} K.~N.,  {Horiuchi} S.,  {Kaplinghat} M.,  {Keeley} R.~E.,
  {Macias} O.,  2020, \mn@doi [\prd] {10.1103/PhysRevD.102.043012}, \href
  {https://ui.adsabs.harvard.edu/abs/2020PhRvD.102d3012A} {102, 043012}

\bibitem[\protect\citeauthoryear{{Abramowski} et~al.,}{{Abramowski}
  et~al.}{2014}]{Abramowski_et_al_2014}
{Abramowski} A.,  et~al., 2014, \mn@doi [\prd] {10.1103/PhysRevD.90.112012},
  \href {https://ui.adsabs.harvard.edu/abs/2014PhRvD..90k2012A} {90, 112012}

\bibitem[\protect\citeauthoryear{Aharonian et~al.}{Aharonian
  et~al.}{2008}]{HESS:2007ora}
Aharonian F.,  et~al., 2008, \mn@doi [Astropart. Phys.]
  {10.1016/j.astropartphys.2007.11.007}, 29, 55

\bibitem[\protect\citeauthoryear{{Albert} et~al.,}{{Albert}
  et~al.}{2017}]{Albert_et_al_2017}
{Albert} A.,  et~al., 2017, \mn@doi [\apj] {10.3847/1538-4357/834/2/110}, \href
  {https://ui.adsabs.harvard.edu/abs/2017ApJ...834..110A} {834, 110}

\bibitem[\protect\citeauthoryear{{Astropy Collaboration} et~al.,}{{Astropy
  Collaboration} et~al.}{2013}]{astropy:2013}
{Astropy Collaboration} et~al., 2013, \mn@doi [\aap]
  {10.1051/0004-6361/201322068}, \href
  {http://adsabs.harvard.edu/abs/2013A%26A...558A..33A} {558, A33}

\bibitem[\protect\citeauthoryear{{Astropy Collaboration} et~al.,}{{Astropy
  Collaboration} et~al.}{2018}]{astropy:2018}
{Astropy Collaboration} et~al., 2018, \mn@doi [\aj] {10.3847/1538-3881/aabc4f},
  \href {https://ui.adsabs.harvard.edu/abs/2018AJ....156..123A} {156, 123}

\bibitem[\protect\citeauthoryear{{Astropy Collaboration} et~al.,}{{Astropy
  Collaboration} et~al.}{2022}]{astropy:2022}
{Astropy Collaboration} et~al., 2022, \mn@doi [apj] {10.3847/1538-4357/ac7c74},
  \href {https://ui.adsabs.harvard.edu/abs/2022ApJ...935..167A} {935, 167}

\bibitem[\protect\citeauthoryear{{Atwood} et~al.,}{{Atwood}
  et~al.}{2009}]{Atwood_et_al_2009}
{Atwood} W.~B.,  et~al., 2009, \mn@doi [\apj] {10.1088/0004-637X/697/2/1071},
  \href {https://ui.adsabs.harvard.edu/\#abs/2009ApJ...697.1071A} {697, 1071}

\bibitem[\protect\citeauthoryear{Belokurov et~al.,}{Belokurov
  et~al.}{2013}]{Belokurov_et_al_2014}
Belokurov V.,  et~al., 2013, \mn@doi [Monthly Notices of the Royal Astronomical
  Society] {10.1093/mnras/stt1862}, 437, 116

\bibitem[\protect\citeauthoryear{{Bertone}, {Hooper}  \& {Silk}}{{Bertone}
  et~al.}{2005}]{Bertone_2005}
{Bertone} G.,  {Hooper} D.,   {Silk} J.,  2005, \mn@doi [\physrep]
  {10.1016/j.physrep.2004.08.031}, \href
  {http://adsabs.harvard.edu/abs/2005PhR...405..279B} {405, 279}

\bibitem[\protect\citeauthoryear{{Bringmann} \& {Weniger}}{{Bringmann} \&
  {Weniger}}{2012}]{Bringmann_and_Weinger_2012}
{Bringmann} T.,  {Weniger} C.,  2012, \mn@doi [Physics of the Dark Universe]
  {10.1016/j.dark.2012.10.005}, \href
  {https://ui.adsabs.harvard.edu/#abs/2012PDU.....1..194B} {1, 194}

\bibitem[\protect\citeauthoryear{{Calore}, {Cholis}, {McCabe}  \&
  {Weniger}}{{Calore} et~al.}{2015}]{Calore_et_al_2015b}
{Calore} F.,  {Cholis} I.,  {McCabe} C.,   {Weniger} C.,  2015, \mn@doi [\prd]
  {10.1103/PhysRevD.91.063003}, \href
  {https://ui.adsabs.harvard.edu/abs/2015PhRvD..91f3003C} {91, 063003}

\bibitem[\protect\citeauthoryear{{Charbonnier} et~al.,}{{Charbonnier}
  et~al.}{2011}]{Charbonnier_et_al_2011}
{Charbonnier} A.,  et~al., 2011, \mn@doi [\mnras]
  {10.1111/j.1365-2966.2011.19387.x}, \href
  {http://adsabs.harvard.edu/abs/2011MNRAS.418.1526C} {418, 1526}

\bibitem[\protect\citeauthoryear{Crocker et~al.,}{Crocker
  et~al.}{2022}]{Crocker_and_Macias_et_al_2022}
Crocker R.~M.,  et~al., 2022, \mn@doi [Nature Astronomy]
  {10.1038/s41550-022-01777-x}

\bibitem[\protect\citeauthoryear{{Diemer}}{{Diemer}}{2018}]{Diemer_2018}
{Diemer} B.,  2018, \mn@doi [\apjs] {10.3847/1538-4365/aaee8c}, \href
  {https://ui.adsabs.harvard.edu/abs/2018ApJS..239...35D} {239, 35}

\bibitem[\protect\citeauthoryear{{Dierickx} \& {Loeb}}{{Dierickx} \&
  {Loeb}}{2017}]{Dierickx_and_Loeb_2017}
{Dierickx} M. I.~P.,  {Loeb} A.,  2017, \mn@doi [\apj]
  {10.3847/1538-4357/836/1/92}, \href
  {https://ui.adsabs.harvard.edu/\#abs/2017ApJ...836...92D} {836, 92}

\bibitem[\protect\citeauthoryear{{Evans}, {Strigari}, {Svenborn}, {Albert},
  {Harding}, {Hooper}, {Linden}  \& {Pace}}{{Evans}
  et~al.}{2023}]{Evans_et_al_2023}
{Evans} A.~J.,  {Strigari} L.~E.,  {Svenborn} O.,  {Albert} A.,  {Harding}
  J.~P.,  {Hooper} D.,  {Linden} T.,   {Pace} A.~B.,  2023, \mn@doi [\mnras]
  {10.1093/mnras/stad2074}, \href
  {https://ui.adsabs.harvard.edu/abs/2023MNRAS.524.4574E} {524, 4574}

\bibitem[\protect\citeauthoryear{{Garrett} \& {D{\=u}da}}{{Garrett} \&
  {D{\=u}da}}{2011}]{Garrett_2011}
{Garrett} K.,  {D{\=u}da} G.,  2011, \mn@doi [Advances in Astronomy]
  {10.1155/2011/968283}, \href
  {http://adsabs.harvard.edu/abs/2011AdAst2011E...8G} {2011, 968283}

\bibitem[\protect\citeauthoryear{Geringer-Sameth, Koushiappas  \&
  Walker}{Geringer-Sameth et~al.}{2015}]{Geringer_Sameth_et_al_2015}
Geringer-Sameth A.,  Koushiappas S.~M.,   Walker M.~G.,  2015, \mn@doi [Phys.
  Rev. D] {10.1103/PhysRevD.91.083535}, 91, 083535

\bibitem[\protect\citeauthoryear{{Geringer-Sameth}, {Koushiappas}, {Walker},
  {Bonnivard}, {Combet}  \& {Maurin}}{{Geringer-Sameth}
  et~al.}{2018}]{Geringer_Sameth_et_al_2018}
{Geringer-Sameth} A.,  {Koushiappas} S.~M.,  {Walker} M.~G.,  {Bonnivard} V.,
  {Combet} C.,   {Maurin} D.,  2018, arXiv e-prints, \href
  {https://ui.adsabs.harvard.edu/\#abs/2018arXiv180708740G} {p.
  arXiv:1807.08740}

\bibitem[\protect\citeauthoryear{{Goodenough} \& {Hooper}}{{Goodenough} \&
  {Hooper}}{2009}]{Goodenough_and_Hooper_2009}
{Goodenough} L.,  {Hooper} D.,  2009, arXiv e-prints, \href
  {https://ui.adsabs.harvard.edu/\#abs/2009arXiv0910.2998G} {p.
  arXiv:0910.2998}

\bibitem[\protect\citeauthoryear{{Gordon} \& {Mac{\'\i}as}}{{Gordon} \&
  {Mac{\'\i}as}}{2013}]{Gordon_and_Macias_2013}
{Gordon} C.,  {Mac{\'\i}as} O.,  2013, \mn@doi [\prd]
  {10.1103/PhysRevD.88.083521}, \href
  {https://ui.adsabs.harvard.edu/abs/2013PhRvD..88h3521G} {88, 083521}

\bibitem[\protect\citeauthoryear{{G{\'o}rski}, {Hivon}, {Banday}, {Wand elt},
  {Hansen}, {Reinecke}  \& {Bartelmann}}{{G{\'o}rski}
  et~al.}{2005}]{Gorski_et_al_2005}
{G{\'o}rski} K.~M.,  {Hivon} E.,  {Banday} A.~J.,  {Wand elt} B.~D.,  {Hansen}
  F.~K.,  {Reinecke} M.,   {Bartelmann} M.,  2005, \mn@doi [\apj]
  {10.1086/427976}, \href
  {https://ui.adsabs.harvard.edu/abs/2005ApJ...622..759G} {622, 759}

\bibitem[\protect\citeauthoryear{{Hernquist}}{{Hernquist}}{1990}]{Hernquist_1990}
{Hernquist} L.,  1990, \mn@doi [\apj] {10.1086/168845}, \href
  {https://ui.adsabs.harvard.edu/abs/1990ApJ...356..359H} {356, 359}

\bibitem[\protect\citeauthoryear{{Ibata}, {Wyse}, {Gilmore}, {Irwin}  \&
  {Suntzeff}}{{Ibata} et~al.}{1997}]{Ibata_et_al_1997}
{Ibata} R.~A.,  {Wyse} R. F.~G.,  {Gilmore} G.,  {Irwin} M.~J.,   {Suntzeff}
  N.~B.,  1997, \mn@doi [\aj] {10.1086/118283}, \href
  {https://ui.adsabs.harvard.edu/abs/1997AJ....113..634I} {113, 634}

\bibitem[\protect\citeauthoryear{Jiang \& Binney}{Jiang \&
  Binney}{2000}]{Jiang_and_Binney_2000}
Jiang I.-G.,  Binney J.,  2000, \mn@doi [Monthly Notices of the Royal
  Astronomical Society] {10.1046/j.1365-8711.2000.03311.x}, 314, 468

\bibitem[\protect\citeauthoryear{{Jungman}, {Kamionkowski}  \&
  {Griest}}{{Jungman} et~al.}{1996}]{Jungman_1996}
{Jungman} G.,  {Kamionkowski} M.,   {Griest} K.,  1996, \mn@doi [\physrep]
  {10.1016/0370-1573(95)00058-5}, \href
  {http://adsabs.harvard.edu/abs/1996PhR...267..195J} {267, 195}

\bibitem[\protect\citeauthoryear{{Kazantzidis}, {{\L}okas}, {Callegari},
  {Mayer}  \& {Moustakas}}{{Kazantzidis} et~al.}{2011}]{Kazantzidis_et_al_2011}
{Kazantzidis} S.,  {{\L}okas} E.~L.,  {Callegari} S.,  {Mayer} L.,
  {Moustakas} L.~A.,  2011, \mn@doi [\apj] {10.1088/0004-637X/726/2/98}, \href
  {https://ui.adsabs.harvard.edu/\#abs/2011ApJ...726...98K} {726, 98}

\bibitem[\protect\citeauthoryear{Kuhlen}{Kuhlen}{2009}]{Kuhlen_2009}
Kuhlen M.,  2009, Advances in Astronomy, 2010

\bibitem[\protect\citeauthoryear{{Law} \& {Majewski}}{{Law} \&
  {Majewski}}{2010}]{Law_and_Majewski_2010}
{Law} D.~R.,  {Majewski} S.~R.,  2010, \mn@doi [\apj]
  {10.1088/0004-637X/714/1/229}, \href
  {https://ui.adsabs.harvard.edu/#abs/2010ApJ...714..229L} {714, 229}

\bibitem[\protect\citeauthoryear{{Law}, {Majewski}, {Skrutskie}  \&
  {Johnston}}{{Law} et~al.}{2004}]{Law_et_al_2004}
{Law} D.~R.,  {Majewski} S.~R.,  {Skrutskie} M.~F.,   {Johnston} K.~V.,  2004,
  in {Prada} F.,  {Martinez Delgado} D.,   {Mahoney} T.~J.,  eds,  Astronomical
  Society of the Pacific Conference Series Vol. 327, Satellites and Tidal
  Streams. p.~239 (\mn@eprint {} {astro-ph/0309567})

\bibitem[\protect\citeauthoryear{Law, Johnston  \& Majewski}{Law
  et~al.}{2005}]{Law_et_al_2005}
Law D.~R.,  Johnston K.~V.,   Majewski S.~R.,  2005, \mn@doi [The Astrophysical
  Journal] {10.1086/426779}, 619, 807

\bibitem[\protect\citeauthoryear{Levenberg}{Levenberg}{1944}]{Levenberg_1944}
Levenberg K.,  1944, Quarterly of applied mathematics, 2, 164

\bibitem[\protect\citeauthoryear{{{\L}okas}, {Kazantzidis}, {Majewski}, {Law},
  {Mayer}  \& {Frinchaboy}}{{{\L}okas} et~al.}{2010}]{Lokas_et_al_2010}
{{\L}okas} E.~L.,  {Kazantzidis} S.,  {Majewski} S.~R.,  {Law} D.~R.,  {Mayer}
  L.,   {Frinchaboy} P.~M.,  2010, \mn@doi [\apj]
  {10.1088/0004-637X/725/2/1516}, \href
  {http://adsabs.harvard.edu/abs/2010ApJ...725.1516L} {725, 1516}

\bibitem[\protect\citeauthoryear{{Macias}, {Gordon}, {Crocker}, {Coleman},
  {Paterson}, {Horiuchi}  \& {Pohl}}{{Macias} et~al.}{2018}]{Macias_et_al_2018}
{Macias} O.,  {Gordon} C.,  {Crocker} R.~M.,  {Coleman} B.,  {Paterson} D.,
  {Horiuchi} S.,   {Pohl} M.,  2018, \mn@doi [Nature Astronomy]
  {10.1038/s41550-018-0414-3}, \href
  {https://ui.adsabs.harvard.edu/\#abs/2018NatAs...2..387M} {2, 387}

\bibitem[\protect\citeauthoryear{{Macias}, {Horiuchi}, {Kaplinghat}, {Gordon},
  {Crocker}  \& {Nataf}}{{Macias} et~al.}{2019}]{Macias_et_al_2019}
{Macias} O.,  {Horiuchi} S.,  {Kaplinghat} M.,  {Gordon} C.,  {Crocker} R.~M.,
   {Nataf} D.~M.,  2019, \mn@doi [\jcap] {10.1088/1475-7516/2019/09/042}, \href
  {https://ui.adsabs.harvard.edu/abs/2019JCAP...09..042M} {2019, 042}

\bibitem[\protect\citeauthoryear{{Majewski}, {Skrutskie}, {Weinberg}  \&
  {Ostheimer}}{{Majewski} et~al.}{2003}]{Majewski_et_al_2003}
{Majewski} S.~R.,  {Skrutskie} M.~F.,  {Weinberg} M.~D.,   {Ostheimer} J.~C.,
  2003, \mn@doi [\apj] {10.1086/379504}, \href
  {https://ui.adsabs.harvard.edu/\#abs/2003ApJ...599.1082M} {599, 1082}

\bibitem[\protect\citeauthoryear{Marquardt}{Marquardt}{1963}]{Marquardt_1963}
Marquardt D.~W.,  1963, Journal of the Society for Industrial and Applied
  Mathematics, 11, 431

\bibitem[\protect\citeauthoryear{{Mazziotta}, {Loparco}, {de Palma}  \&
  {Giglietto}}{{Mazziotta} et~al.}{2012}]{Mazziotta_et_al_2012}
{Mazziotta} M.~N.,  {Loparco} F.,  {de Palma} F.,   {Giglietto} N.,  2012,
  \mn@doi [Astroparticle Physics] {10.1016/j.astropartphys.2012.07.005}, \href
  {https://ui.adsabs.harvard.edu/abs/2012APh....37...26M} {37, 26}

\bibitem[\protect\citeauthoryear{{Navarro}, {Frenk}  \& {White}}{{Navarro}
  et~al.}{1997}]{Navarro_et_al_1997}
{Navarro} J.~F.,  {Frenk} C.~S.,   {White} S. D.~M.,  1997, \mn@doi [\apj]
  {10.1086/304888}, \href
  {https://ui.adsabs.harvard.edu/\#abs/1997ApJ...490..493N} {490, 493}

\bibitem[\protect\citeauthoryear{Newberg \& Carlin}{Newberg \&
  Carlin}{2016}]{Newberg_and_Carlin_2016}
Newberg H.~J.,  Carlin J.~L.,  eds, 2016, {Tidal Streams in the Local Group and
  Beyond}  Tidal Streams in the Local Group and Beyond Vol. 420,
  \mn@doi{10.1007/978-3-319-19336-6.
}

\bibitem[\protect\citeauthoryear{Pohl, Macias, Coleman  \& Gordon}{Pohl
  et~al.}{2022}]{Pohl:2022nnd}
Pohl M.,  Macias O.,  Coleman P.,   Gordon C.,  2022, \mn@doi [Astrophys. J.]
  {10.3847/1538-4357/ac6032}, 929, 136

\bibitem[\protect\citeauthoryear{{Power}, {Navarro}, {Jenkins}, {Frenk},
  {White}, {Springel}, {Stadel}  \& {Quinn}}{{Power}
  et~al.}{2003}]{Power_et_al_2003}
{Power} C.,  {Navarro} J.~F.,  {Jenkins} A.,  {Frenk} C.~S.,  {White} S.~D.~M.,
   {Springel} V.,  {Stadel} J.,   {Quinn} T.,  2003, \mn@doi [\mnras]
  {10.1046/j.1365-8711.2003.05925.x}, \href
  {https://ui.adsabs.harvard.edu/abs/2003MNRAS.338...14P} {338, 14}

\bibitem[\protect\citeauthoryear{Rico}{Rico}{2020}]{Rico_et_al_2020}
Rico J.,  2020, \mn@doi [Galaxies] {10.3390/galaxies8010025}, 8

\bibitem[\protect\citeauthoryear{{Stoehr}, {White}, {Springel}, {Tormen}  \&
  {Yoshida}}{{Stoehr} et~al.}{2003}]{Stoehr_et_al_2003}
{Stoehr} F.,  {White} S. D.~M.,  {Springel} V.,  {Tormen} G.,   {Yoshida} N.,
  2003, \mn@doi [\mnras] {10.1046/j.1365-2966.2003.07052.x}, \href
  {https://ui.adsabs.harvard.edu/abs/2003MNRAS.345.1313S} {345, 1313}

\bibitem[\protect\citeauthoryear{{Tepper-Garc{\'\i}a} \&
  {Bland-Hawthorn}}{{Tepper-Garc{\'\i}a} \&
  {Bland-Hawthorn}}{2018}]{Tepper_Garcia_and_Bland_Hawthorn_2018}
{Tepper-Garc{\'\i}a} T.,  {Bland-Hawthorn} J.,  2018, \mn@doi [\mnras]
  {10.1093/mnras/sty1359}, \href
  {https://ui.adsabs.harvard.edu/\#abs/2018MNRAS.478.5263T} {478, 5263}

\bibitem[\protect\citeauthoryear{{Teyssier}}{{Teyssier}}{2002}]{Teyssier_2002}
{Teyssier} R.,  2002, \mn@doi [\aap] {10.1051/0004-6361:20011817}, \href
  {https://ui.adsabs.harvard.edu/abs/2002A&A...385..337T} {385, 337}

\bibitem[\protect\citeauthoryear{{Viana} et~al.,}{{Viana}
  et~al.}{2012}]{Viana_et_al_2012}
{Viana} A.,  et~al., 2012, \mn@doi [\apj] {10.1088/0004-637X/746/1/77}, \href
  {https://ui.adsabs.harvard.edu/abs/2012ApJ...746...77V} {746, 77}

\bibitem[\protect\citeauthoryear{Virtanen et~al.,}{Virtanen
  et~al.}{2020}]{Virtanen_et_al_2020}
Virtanen P.,  et~al., 2020, \mn@doi [Nature Methods]
  {10.1038/s41592-019-0686-2}, \href {https://rdcu.be/b08Wh} {17, 261}

\makeatother
\end{thebibliography}


\appendix
\section{Equivalence of the J-factor definition}
\label{sec:Charbonnier_equivalence_appendix}
In section~\ref{sec:J_factor_calculation_method}, equation~\ref{eq:EQ6} defined the J-factor value for a simulated dark matter particle distribution occupying a given volume element. This definition can be shown to be equivalent to the J-factor definition defined in equation 5 of \citet{Charbonnier_et_al_2011}, excepting the factor of $1/4\pi$ introduced in this study to account for the surface area of the flux sphere for each simulation particle involved in the summation detailed in equation~\ref{eq:EQ6}. Starting with equation 5 of \citet{Charbonnier_et_al_2011}:
\begin{equation}
    J=\int_{\Delta \Omega}\int \rho^2_{DM} (l,\Omega)\text{d}l\text{d}\Omega
    \label{Charbonnier_equation_5}
\end{equation}
we first make the change of notation $l=r$ to convert this equation to the notation of this study. It can then be shown that 
\begin{multline}
    J=\int_{\Delta \Omega}\int \rho^2_{DM} (r,\Omega)\text{d}r\text{d}\Omega = \int \int \rho_{DM}(r,\Omega)^2/(r^2) r^2\ \text{d}r\text{d}\Omega \\
    \label{EQA2}
\end{multline}
The right-hand side of equation~\ref{EQA2} is equivalent to an integral over a volume element $b$ of radial length $r$ and angular size $\Delta \Omega$:
\begin{equation}
    J_b = \int \int \rho_{DM}(r,\Omega)^2/(r^2) r^2\ \text{d}r\text{d}\Omega \\= \int_{V_b} \rho_{DM}^2/(r^2)\ \text{d}V
    \label{EQA3}
\end{equation}
Finally, adding the additional factor of $1/4\pi$ to account for the surface area of the flux sphere for each particle, we reach the J-factor definition defined in the left-hand side of equation~\ref{eq:EQ6}:
\begin{equation}
    J_b = \frac{1}{4\pi}\int_{V_b} \rho_{DM}^2/(r^2)\ \text{d}V=\int_{V_b} \rho_{DM}^2/(4\pi r^2)\ \text{d}V
\end{equation}
Accordingly, the J-factor definition defined in equation 5 of \citet{Charbonnier_et_al_2011} is equivalent to the J-factor definition of equation~\ref{eq:EQ6} excepting the differing factor of $4\pi$, when defined for a given volume element $b$ of angular size $\Delta \Omega$ and radial length $r$.
\section{The two-dimensional J-factor distribution in physical units}
\label{sec:2D_physical_unit_distributions_appendix}
\begin{figure}
    \centering 
    \includegraphics[width=\columnwidth]{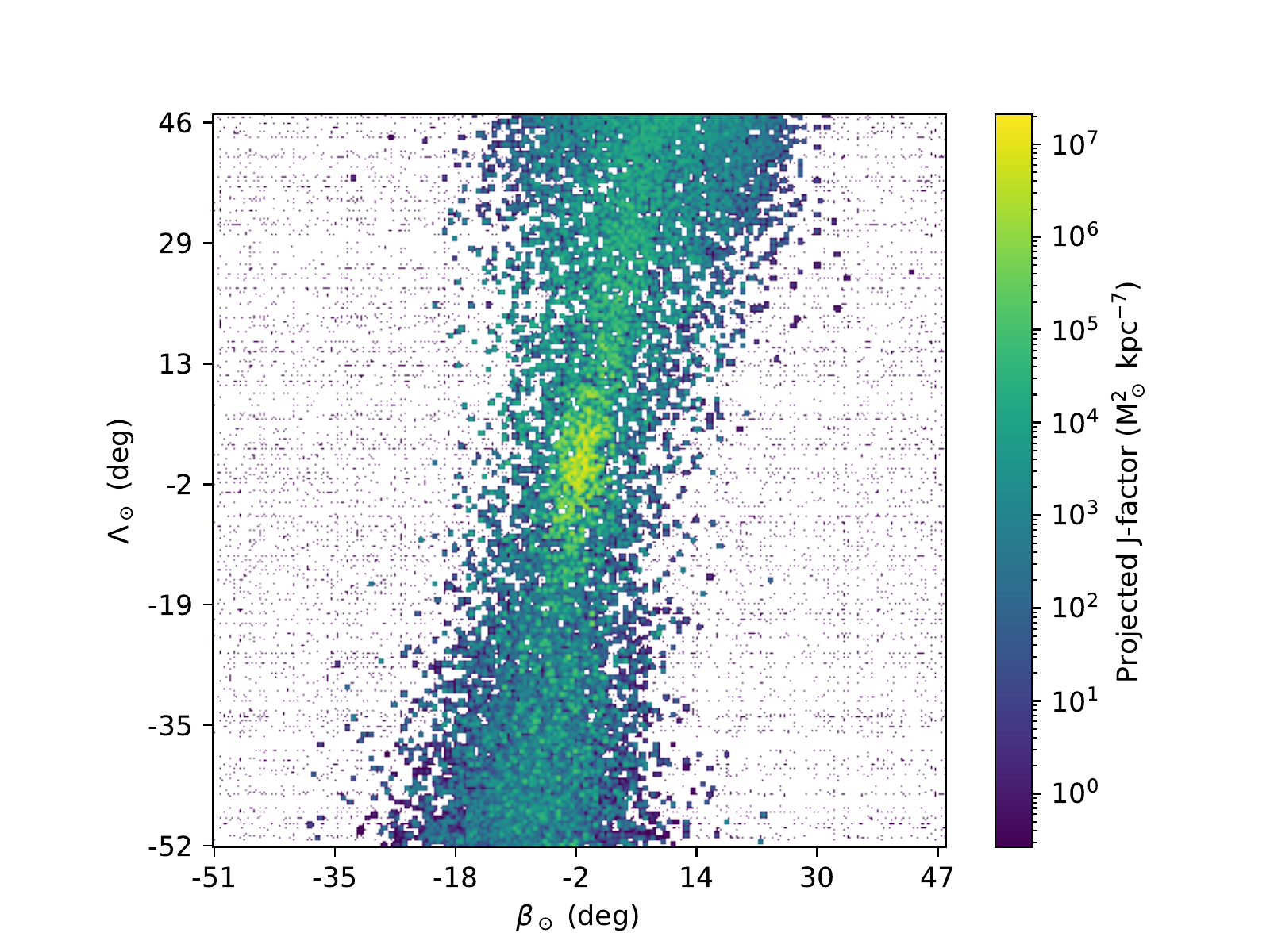}
    \caption{The projected J-factor distribution centred on the location of the simulated Sagittarius Dwarf, again in the coordinate system of \citet{Belokurov_et_al_2014}. This distribution was calculated from the DM J-factor distribution depicted in Figure~\ref{fig:J_factor_DM_ROI_angular_units}, which had the units of M$_\odot^2$ kpc$^{-5}$ deg$^{-2}$, by division of each pixel by the factor $C_b$ detailed in equation~\ref{eq:C}. The adopted pixel size in this figure remains $\alpha^2 = 0.21$ square degrees.}
    \label{fig:J_factor_DM_ROI_physical_units}
\end{figure}

Section~\ref{sec:J_factor_calculation_method} details the production of the projected J-factor distribution of Sgr. As discussed in section~\ref{sec:J_factor_calculation_method}, the 2D projected J-factor distribution (depicted in Figure~\ref{fig:J_factor_DM_ROI_angular_units}) was converted from the angular units of M$_\odot^2$ kpc$^{-5}$ deg$^{-2}$ to the physical units of M$_\odot^2$ kpc$^{-7}$ through the division of each pixel $b$ by the factor $C_b$ detailed in equation~\ref{eq:C}. The resulting projected J-factor distribution is displayed in Figure~\ref{fig:J_factor_DM_ROI_physical_units}.
\section{Scaled 1D J-factor profiles}
\label{sec:Scaled_1D_profiles_appendix}
\begin{figure}
    \includegraphics[width=\columnwidth]{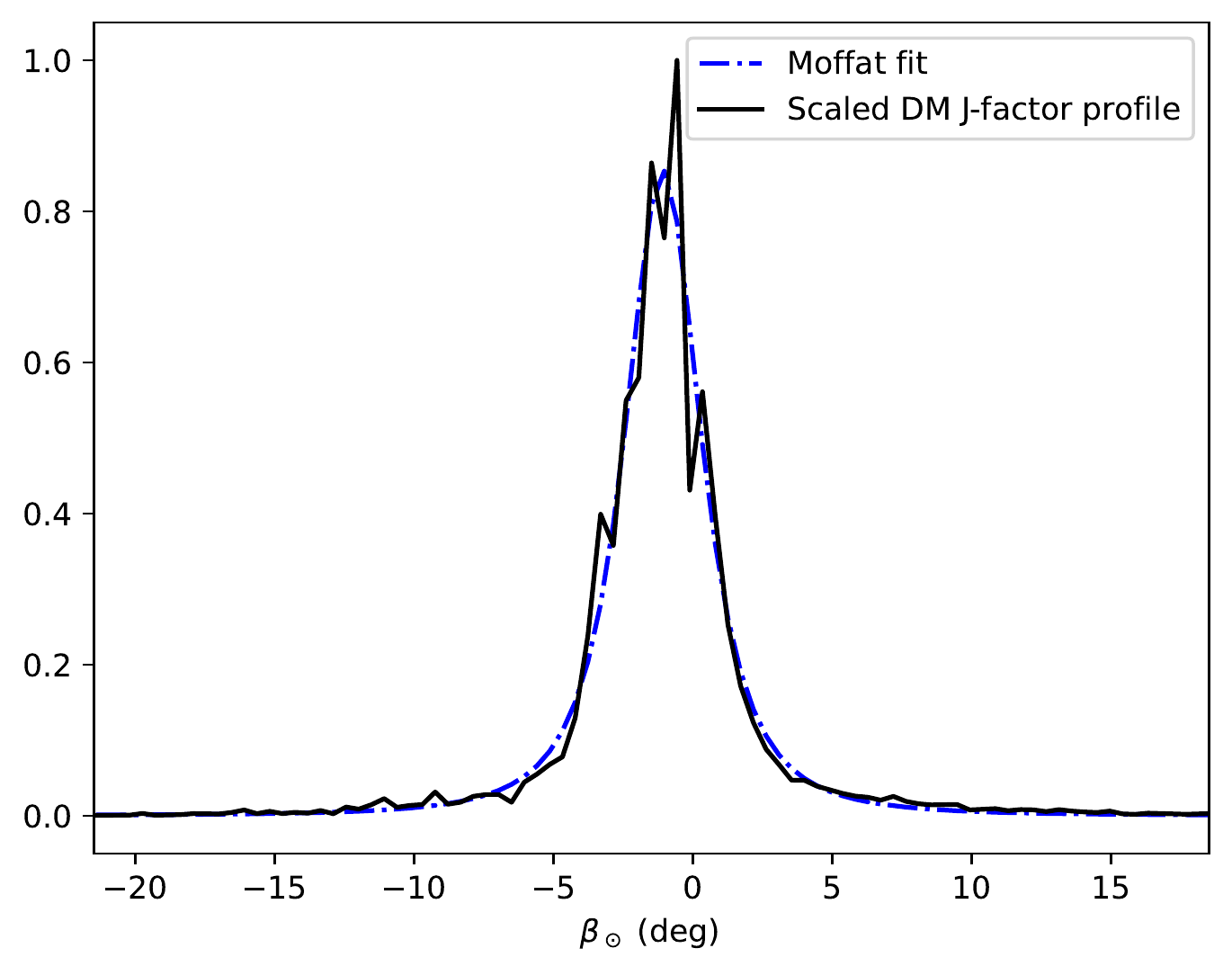}
    \caption{The dark matter $\beta_\odot$ J-factor profile (in the angular units of M$_\odot^2$ kpc$^{-5}$ deg$^{-1}$), scaled such that the maximum value is equal to $1.0$. Also shown is the fitted Moffat distribution. The FWHM of this fitted distribution is $3.7^\circ$. The adopted $\beta_\odot$ bin size remains $0.458^\circ$.}
    \label{fig:DM_J_factor_Beta_profile_normalized}
\end{figure}

\begin{figure}
    \includegraphics[width=\columnwidth]{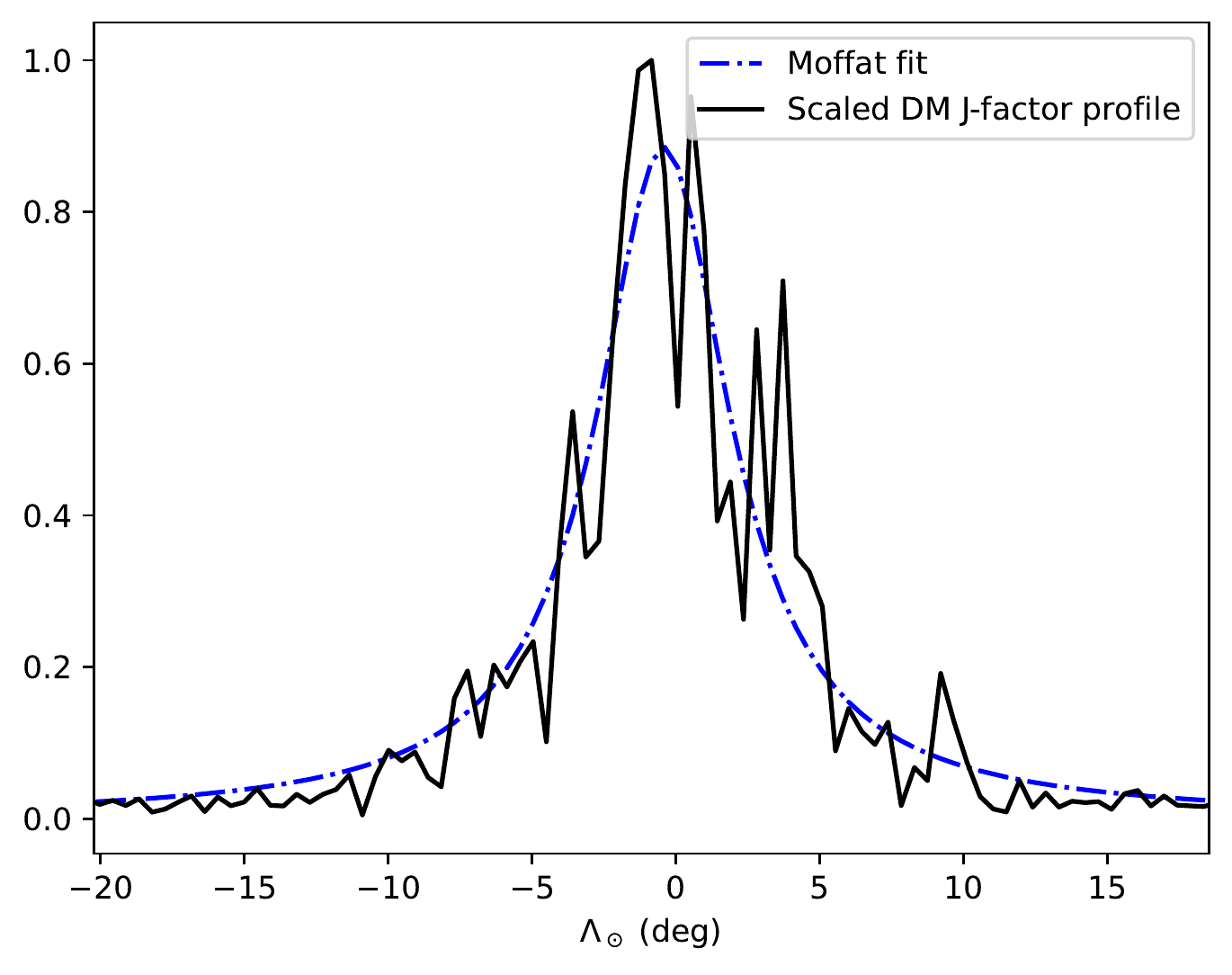}
    \caption{The dark matter $\Lambda_\odot$ J-factor profile (in the angular units of M$_\odot^2$ kpc$^{-5}$ deg$^{-1}$), scaled such that the maximum value is equal to $1.0$. The adopted $\Lambda_\odot$ bin size remains $0.458^\circ$. The FWHM of the fitted Moffat distribution is $4.9^\circ$.}
    \label{fig:DM_J_factor_Lambda_profile_normalized}
\end{figure}

In section~\ref{sec:1D_J_factor_profiles_results}, 1D J-factor profiles for Sgr were calculated from the 2D J-factor distribution to illustrate morphological features of the J-factor distribution and inform observational searches. The scaled versions of these profiles were fitted with functions (either a Voigt, Moffat, Gaussian or Gaussian with constant vertical offset) through minimisation of the optimization function detailed in equation~\ref{eq:Opt_function}. The parameters of the best fitting functions are described in Table~\ref{tab:Fits_summary_table}. Only the profiles with angular units (of M$_\odot^2$ kpc$^{-5}$ deg$^{-1}$) were scaled to avoid any small scale variation caused by variation in the applied divisor $C_b$ between pixels on profiles with the physical units (namely, M$_\odot^2$ kpc$^{-7}$ deg).
\\[10pt]
Figure~\ref{fig:DM_J_factor_Beta_profile_normalized} illustrates the scaled J-factor profile as a function of $\beta_\odot$ whilst Figure~\ref{fig:DM_J_factor_Lambda_profile_normalized} illustrates the scaled J-factor profile as a function of $\Lambda_\odot$. Both of these figures also display the fitted Moffat functions (defined in equation~\ref{eq:Moffat}).
\\[10pt]
%
%
%
%
\bsp	
\label{lastpage}
\end{document}